\documentclass[paper]{JHEP3} 

\JHEPspecialurl{http://jhep.sissa.it/JOURNAL/JHEP3.tar.gz}

\usepackage{epsfig,multicol}

\newcommand\fverb{\setbox\pippobox=\hbox\bgroup\verb}
\newcommand\fverbdo{\egroup\medskip\noindent%
			\fbox{\unhbox\pippobox}\ }
\newcommand\fverbit{\egroup\item[\fbox{\unhbox\pippobox}]}
\newcommand {\beq}{\begin{equation}}
\newcommand {\eeq}{\end{equation}}
\newcommand {\beqa}{\begin{eqnarray}}
\newcommand {\eeqa}{\end{eqnarray}}

\newcommand {\tr}{{\rm tr\,}}
\newcommand {\Tr}{\mbox{Tr\,}}

\newcommand {\ee}{\mbox{e}}

\newcommand {\dd}{\mbox{d}}

\newcommand {\del}{\partial}

\newbox\pippobox

\title{Gaussian expansion analysis of a matrix model\\
with the spontaneous breakdown of rotational symmetry}

\author{Jun Nishimura\\
High Energy Accelerator Research Organization (KEK)\\
1-1 Oho, Tsukuba 305-0801, Japan\\
E-mail: \email{jnishi@post.kek.jp}}
\author{Toshiyuki Okubo\\
Department of Physics, Nagoya University\\
Furo-cho, Chikusa-ku, Nagoya 464-8602, Japan\\
E-mail: \email{okubo@eken.phys.nagoya-u.ac.jp}}
\author{Fumihiko Sugino\\
Okayama Institute for Quantum Physics\\
Kyoyama 1-9-1, Okayama 700-0015, Japan\\
E-mail: \email{fumihiko\_sugino@pref.okayama.jp}}

\preprint{KEK-TH-1003\\
DPNU-04-23\\
OIQP-04-08\\
\hepth{0412194}}	

\abstract{
Recently the gaussian expansion method
has been applied to investigate
the dynamical generation of 4d space-time
in the IIB matrix model,
which is a conjectured nonperturbative definition
of type IIB superstring theory in 10 dimensions.
Evidence for such a phenomenon, which is associated
with the spontaneous breaking of the SO(10) symmetry down to SO(4), 
has been obtained up to the 7-th order calculations.
Here we apply the same method to 
a simplified model, which is expected to exhibit an analogous
spontaneous symmetry breaking via the same mechanism as conjectured 
for the IIB matrix model.
The results up to the 9-th order demonstrate a clear convergence,
which allows us to unambiguously identify the actual
symmetry breaking pattern by comparing the free energy
of possible vacua and to calculate the extent of ``space-time''
in each direction.
}


\begin{document} 


\section{Introduction}

It has been long considered that matrix models may be useful as a 
nonperturbative formulation of string theory, and hence
play an important role similar to the lattice formulation in
quantum field theory.
Indeed matrix models have been quite successful in formulating 
non-critical string theory, and after the development of the notions
such as ``string duality'' and ``D-branes'',
the idea has been extended also to critical strings.
The IIB matrix model \cite{IKKT} is one of such proposals,
which is conjectured to be a nonperturbative definition
of type IIB superstring theory in 10 dimensions.
It is a supersymmetric matrix model, which can be formally obtained
by the zero-volume limit
of 10d SU($N$) super Yang-Mills theory.

In this model the space-time is represented
by the eigenvalue distribution of ten bosonic matrices \cite{AIKKT}.
If the distribution collapses dynamically
to a four-dimensional hypersurface, 
which in particular requires
the SO(10) symmetry of the model to be spontaneously broken,
we may naturally understand the dimensionality of our space-time 
as a result of the nonperturbative dynamics of superstring theory.
In ref.~\cite{Nishimura:2001sx} the first evidence for the above scenario
has been obtained by calculating the free energy of space-time
with various dimensionality using the gaussian expansion method up to
the 3rd order.  Higher-order calculations~\cite{KKKMS,KKKS} as well as
the tests of the method itself 
in simpler models~\cite{NOS1,NOS2} have
strengthened the conclusion considerably.  
Refs.\ \cite{0307007} provide another evidence for the emergence
of four-dimensional space-time based on perturbative calculations around
fuzzy-sphere like solutions.

While these results are certainly encouraging, it is desirable to
understand the mechanism for the spontaneous symmetry breaking (SSB) of 
the rotational symmetry.
In refs.\ \cite{NV} it has been pointed out that
the phase of the fermion determinant
favors lower dimensional configurations since the phase becomes
stationary around such configurations
\footnote{See refs.\ \cite{AIKKT,Burda:2000mn,Vernizzi:2002mu} for
discussions on other possible mechanisms.}.
Indeed Monte Carlo simulations show
that SSB does {\em not} occur in various models {\em without}
such a phase factor \cite{HNT,4dSSB,branched}.
Unfortunately including the effects of the phase in Monte Carlo simulation 
is technically difficult due to the so-called sign problem,
but a new method \cite{sign}, which is tested in Random Matrix Theory \cite{RMT},
was able to produce some preliminary results, which look promising.
In ref.\ \cite{exact} a simple matrix model which 
realizes the above mechanism has been proposed.
The model contains $N_{\rm f}$ flavors of Weyl fermion in the fundamental 
representation of SU($N$), which yield a complex fermion determinant,
and the large $N$ limit is taken with the ratio $r = N_{\rm f}/N$ being fixed.
The model can be solved exactly at infinitesimally small $r$,
and the SO(4) symmetry is shown to be broken down to SO(3).

In this paper we study this model at {\em finite} $r$ by
the gaussian expansion method.
Since the model is much simpler than the IIB matrix model, 
we can perform calculations up to the 9-th order with reasonable efforts.
The results demonstrate a clear convergence for $r \lesssim 2$,
which allows us to unambiguously identify the
symmetry breaking pattern and to calculate the extent of ``space-time''
in each direction. 

In fact it turns out that the SO(4) symmetry is broken 
down to SO(2) at finite $r$.
However, at small $r$ 
we reproduce the free energy as well as the extent 
of ``space-time'' in each direction obtained in ref.\ \cite{exact},
which implies that the SO(3) symmetry is realized
asymptotically as $r$ approaches zero.
In the large $r$ region, on the other hand, 
the extent of ``space-time'' in two directions, 
in which the SO(2) symmetry is realized,
becomes much larger than the remaining two directions.
This behavior can be understood from the viewpoint of refs.\ \cite{NV}
since the phase of the fermion
determinant becomes stationary for two-dimensional configurations, 
and increasing $r$ tends to amplify the effect of the phase.
Thus our results nicely demonstrate the proposed mechanism
for the dynamical generation of space-time in the IIB matrix model.

The rest of this paper is organized as follows.
In Section \ref{model} we define the model
and review the known results.
In Section \ref{GEM} we explain how to apply
the gaussian expansion method to the model.
In Section \ref{results} we present our results.
Section \ref{summary} is devoted to a summary and discussions.
The details of our calculations are given in the Appendix.

\section{The model}
\label{model}

The model we study in this paper is defined by the partition function
\cite{exact}
\beqa
Z &=&  \int \dd A \, \dd \psi \, \dd \bar{\psi} \, 
\ee^{-(S_{\rm b} + S_{\rm f})}  \ ,
\label{original_model} \\
S_{\rm b} &=& \frac{1}{2} \, N \, \tr (A_{\mu})^2  \ ,
\label{actionB}
\\ 
S_{\rm f} &=& - \, N \bar{\psi}_\alpha^{f} (\Gamma_\mu)_{\alpha\beta}
A_\mu \psi_\beta^{f} \ ,
\label{actionPSI}
\eeqa
where $A_{\mu}$ ($\mu = 1,\cdots , 4$) are $N\times N$ traceless
\footnote{The tracelessness condition was not imposed
in the original paper \cite{exact}. While this condition does not affect
the large $N$ limit of the model, it simplifies our calculation drastically;
see footnote \ref{tracepart}.
}
hermitian matrices and $\bar{\psi}_\alpha^f$, 
$\psi_\alpha^f$ ($\alpha = 1,2$;
$f=1 , \cdots , N_{\rm f} $) are $N$-dimensional row and column vectors, 
respectively, making the system SU($N$) invariant.
The integration measure for $A_\mu$ is given by
\beq
 \dd A = \prod_{a=1}^{N^2-1} \prod_{\mu = 1}^{4}
\frac{\dd A_\mu ^a}{\sqrt{2 \pi}} \ ,
\eeq
where $A_\mu ^a$ is the coefficient in the expansion
$A_\mu = \sum_{a=1}^{N^2-1} A_\mu ^a \, T^a $
with respect to the SU($N$) generators $T^a$ ($a=1,\cdots ,(N^2-1)$) 
normalized as $\tr (T^a T^b) = \frac{1}{2} \delta ^{ab}$.
The integration measure for the fermions is given by
\beq
\dd \psi \, \dd \bar{\psi}
= \prod_{f=1}^{N_{\rm f}} \prod_{i=1}^N \prod_{\alpha=1}^2
\dd \psi_{\alpha} ^{fi} \, 
\dd \bar{\psi}_{\alpha} ^{fi} \ .
\eeq

The system has an SO($4$) symmetry,
under which $A_\mu$ transforms as a vector,
and $\psi_\alpha^f$ and $\bar{\psi}_\alpha^f$
transform as Weyl spinors.
The $2\times 2$ matrices $\Gamma_\mu$ 
are the gamma matrices after the Weyl projection, and an explicit form
is given for instance by
  \begin{eqnarray}
   \Gamma_{1} =  
\left( \begin{array}{cc} 0 & 1 \\ 1 & 0 
   \end{array} \right), \hspace{2mm}
   \Gamma_{2} =  
\left( \begin{array}{cc} 0 & -i \\ i & 0
   \end{array} \right), \hspace{2mm}
   \Gamma_{3} =  
\left( \begin{array}{cc} 1 & 0 \\ 0 & -1
   \end{array} \right), \hspace{2mm}
   \Gamma_{4} = 
\left( \begin{array}{cc} i & 0 \\ 0 & i
   \end{array} \right) \ . 
\label{gamma-matrices}
  \end{eqnarray}
The fermionic part of the model can be thought of 
as the zero-volume limit of the system of 
Weyl fermions in four dimensions interacting with a background gauge field
via fundamental coupling.

We take the large $N$ limit keeping the ratio
$r \equiv N_{\rm f} / N $ fixed (Veneziano limit). 
In order to discuss the SSB of the SO(4) symmetry
in that limit, 
we consider the ``moment of inertia tensor'' \cite{AIKKT,HNT}
\beq
T_{\mu\nu} = \frac{1}{N} \tr( A_{\mu} A_{\nu} )  \ ,
\label{def-inertia}
\eeq
which is a $4 \times 4$ real symmetric tensor,
and denote its eigenvalues as $\{ \lambda_i \, ; \, i=1,\cdots , 4\}$
with the specified order
\beq
\lambda_1 \ge \lambda_2 \ge  \lambda_3 \ge \lambda_4 \ .
\eeq
If the vacuum expectation values (VEVs)
$\langle \lambda_i \rangle$ ($i=1,\cdots , 4$)
do not agree in the large $N$ limit, 
we may conclude that the SSB occurs.
Thus $\langle \lambda_i \rangle$ plays the role of
an order parameter.
In the present model the sum of the VEVs is given by
\beq
\sum _{i=1}^4 \langle \lambda_i \rangle = 
\sum_{\mu=1}^4 \left\langle 
\frac{1}{N} \tr( A_{\mu} )^2  \right\rangle = 
4 \left( 1 - \frac{1}{N^2} \right)+ 2 \, r 
\eeq
for arbitrary $N$ and $r$ due to a ``virial theorem'' \cite{exact}.

At infinitesimally small $r$ the VEVs can be obtained in the large $N$
limit as \cite{exact}
\beqa
\langle \lambda_1 \rangle = 
\langle \lambda_2 \rangle = 
\langle \lambda_3 \rangle &=& 
1 +  r + {\rm o}(r)  \ ,  \nonumber \\
\langle \lambda_4 \rangle &=& 1 - r + {\rm o}(r)  \ ,
\label{SSBresult}
\eeqa
which means that the SO(4) symmetry is spontaneously broken down to SO(3).
The SSB is associated with the formation of a condensate
$\langle \bar{\psi}_\alpha ^f \psi_\alpha ^f  \rangle$, which
is invariant under SO(3), but not under the full SO(4) transformation.

An important feature of the model that is relevant to the SSB
is that the fermion determinant $\det {\cal D}$,
where ${\cal D}$ is a $2 N \times 2 N$ matrix given by
${\cal D} = \Gamma_\mu A_\mu$, is complex in general.
If one replaces the fermion determinant by its absolute value,
the same analysis at infinitesimal $r$ leads to 
an SO(4) symmetric result \cite{exact}.
Thus the SSB of the original model occurs precisely due to the phase
of the fermion determinant.

At large $r$ the effect of the phase is amplified, and we may
expect that the configurations for which the phase is stationary
dominate the path integral.
Analogously to the situation in the IIB matrix model \cite{NV},
the phase becomes stationary for 2-dimensional configurations
in the present model.
Therefore we anticipate the emergence of 2-dimensional
``space-time'' 
($ \lambda_1 $, 
$ \lambda_2 $ $\gg$ 
$ \lambda_3 $, 
$ \lambda_4 $)
as $r$ increases
\footnote{
Note that the phase of the fermion determinant is invariant under
the scale transformation $A_\mu \mapsto \alpha A_\mu$.
Therefore it is only the {\em ratio} of the eigenvalues that matters
for the stationarity.}.


\section{The gaussian expansion method}
\label{GEM}

We are going to obtain results for the model (\ref{original_model})
at finite $r$ using the gaussian expansion method.
The method has a long history, and the original idea appeared 
already around 1980 
in the context of solving quantum mechanical systems 
\cite{conv,Stevenson:1981vj}, where
the expansion was shown to be convergent 
in some concrete examples \cite{exact_conv}.
The method proved useful also in field theories \cite{GEM_field}
in various contexts.
Applications to superstring/M theories using
their matrix model formulations have been advocated 
by Kabat and Lifschytz \cite{Kabat:2000hp},
and the subsequent series of papers \cite{blackholes} revealed
interesting blackhole thermodynamics.
Applications to simplified versions of the IIB matrix model
were initiated in refs.\ \cite{Gauss_simpleIIB}.
An earlier application to random matrix models can be found
in ref.\ \cite{EL}.

Similarly to the case of the IIB matrix model \cite{Nishimura:2001sx,KKKMS,KKKS}, 
let us introduce the gaussian action
\footnote{
\label{tracepart}
If we did not impose the tracelessness condition on $A_\mu$,
we would need to consider a linear term such as 
$S_{\rm lin} = N \sum_\mu h_\mu \tr A_\mu$
in the gaussian action (\ref{defS0}).
The gaussian expansion method can be extended 
to such a case \cite{NOS2}, but the calculation will be more involved.}
\beq
S_0  = 
\frac12 \,  N \sum_{\mu=1}^4  t_{\mu} \, \tr (A_{\mu}) ^2
+ N \sum_{f=1}^{N_{\rm f}}\sum_{\alpha, \beta =1}^2 
{\cal A}_{\alpha \beta} \, 
\bar{\psi}_{\alpha}^{f} \,   \psi_{\beta}^{f} \ , 
\label{defS0}
\eeq
which breaks the SO(4) symmetry. 
The $2\times 2$ complex matrix ${\cal A}$ can be expanded 
in terms of gamma matrices as
\beq
{\cal A} = \sum_{\mu =1}^4 u_{\mu}\Gamma_{\mu}
\label{defA}
\eeq
using 4 complex parameters $u_{\mu}$.
Then we consider the action
\beq
S_{\rm GEM} (t, u ; \lambda)
= \, \frac{1}{\lambda} \, \Bigl[ \Bigl\{ S_{0} + 
\lambda (S_{\rm b} - S_{0} ) 
\Bigr\}  + S_{\rm f}  \Bigr] \  ,
\label{GEM_action}
\eeq
which reduces to the original action for $\lambda = 1$.
The gaussian expansion amounts
to calculating various quantities as an expansion with respect
to $\lambda$ up to some finite order and setting $\lambda = 1$ eventually.
As we will discuss shortly, the free parameters $t_\mu$ and $u_\mu$
in the gaussian action $S_0$ play a crucial role in the method.

In fact the gaussian expansion can be viewed as a loop expansion
with the ``classical action'' $(S_0 + S_{\rm f})$ 
and the ``one-loop counterterms'' $(S_{\rm b} - S_0)$.
This becomes clear upon rescaling $A_\mu$ and $\psi$ as
$A_\mu \mapsto \lambda \, A_\mu$
$\psi_\alpha^{f} \mapsto \sqrt{\lambda} \, \psi_\alpha^{f}$,
so that the partition function takes the form
\beqa
Z &=& \int \dd A \, \dd \psi \, \dd \bar{\psi} \, 
e^{ - ( S_{\rm cl} + S_{\rm c.t.}) } \ ,  \\
S_{\rm cl} (t,u) &=& S_0 + \sqrt{\lambda} \, S_{\rm f}
   ~~~,~~~
S_{\rm c.t.} (t,u) =  \lambda \, (S_{\rm b} - S_0 )  \ .
\eeqa
In actual calculations 
the ``one-loop counter terms'' can be 
incorporated easily by exploiting the relation
\beq
S_{\rm cl} (t, u)  + 
S_{\rm c.t.} (t,u ) 
= S_{\rm cl} (t + \lambda (1-t), 
u- \lambda  u  ) \ .
\eeq

As an example, let us consider evaluating the free energy 
$F = - \frac{1}{N^2}\ln Z$
by the gaussian expansion method.
We first calculate the free energy ${\cal F}(t,u)$ for
the ``classical action'' $S_{\rm cl} (t,u)$
defined by
\beq
\exp[-N^2 {\cal F}(t,u)] = 
\int \dd A \, \dd\psi \, \dd\bar{\psi} \, 
 e^{- S_{\rm cl}(t,u)  } \ .
\label{calF}
\eeq
This can be done by ordinary Feynman diagrammatic
calculations, where the use of Schwinger-Dyson (SD) equations
reduces the number of diagrams considerably \cite{KKKMS}.
Suppose we obtain the result up to the $K$-th order as
\beq
{\cal F}_K(t,u)= \sum_{k=0}^K  {\widetilde {\cal F}}_k (t,u) \, \lambda^k
\ .
\eeq
We shift the arguments, and obtain the new coefficients 
${\tilde  F}_k (t,u)$ in the expansion
\beq
{\cal F}_K(t+\lambda (1-t), u -\lambda u)
= \sum_{k=0}^K  {\tilde  F}_k (t,u)\,  \lambda^k
+ {\rm O}(\lambda ^{K+1})\ .
\eeq
Then the free energy for the original model can be evaluated as 
\beq
F_K(t,u) 
= \sum_{k=0}^K  {\tilde  F}_k (t,u) \ .
\eeq


The result of such calculations
depends on the free parameters in the gaussian action.
However, in various models \cite{KKKMS,NOS1,NOS2}
a ``plateau'' region, in which the result becomes almost constant,
was found to develop in the parameter space
as one goes to higher orders of the expansion.
Moreover it turned out that the height of the plateau agrees 
very accurately with the correct value obtained by some other method.
Therefore it is reasonable to expect that the method works
in general if one can identify a plateau in the parameter space.
In old literature the free parameters were determined in such a way 
that the result becomes most insensitive
to the change of the parameters \cite{Stevenson:1981vj},
but it is really the formation of a plateau 
that ensures the validity of the method
as has been first recognized in ref.\ \cite{KKKMS}.


Identification of a plateau becomes a non-trivial issue when there are
many parameters in the gaussian action.
The histogram prescription \cite{NOS1,NOS2}, 
which works nicely when there are only one or two real parameters,
does not seem to work when the number exceeds three.
(Note that we have 4 real and 4 complex parameters
in the present case.)
We have also attempted a Monte Carlo simulation
{\em in the parameter space} to search for a plateau but with little success.
We therefore use the prescription adopted
for the IIB matrix model.
First we solve the ``self-consistency equations''
\beqa
\frac{\del}{\del t_\mu} 
F_K(t,u)  &=& 0   \ , \nonumber \\
\frac{\del}{\del u_\mu} 
F_K(t,u)  &=& 0  \ .
\label{self-consistency_eq}
\eeqa
Typically we obtain many solutions 
as we go to higher orders. If we observe that solutions concentrate
in some region of the parameter space, we consider it as an indication
of the plateau formation.


Although the number of parameters 
is much less than 
that (10 real and 120 complex parameters) in the IIB matrix model,
it is still difficult to obtain all the solutions
of the self-consistency equations at high orders.
As is done in the case of the IIB matrix model
\cite{Nishimura:2001sx,KKKMS,KKKS},
we search for solutions
assuming that some subgroup of the full rotational symmetry is preserved.
Here we consider the following Ans\"atze.
For each case the independent parameters
will be 2 real and 1 complex numbers.

\paragraph{SO(3) Ansatz :}
We assume SO(3) symmetry in the $x_2$, $x_3$, $x_4$ directions.
Then the parameters are restricted to 
\beq
t_2=t_3=t_4 (\equiv \tilde{t}), \qquad u_2=u_3=u_4=0 \ .
\eeq

\paragraph{SO(2) Ansatz :}
We assume SO(2) symmetry in the $x_3$, $x_4$ directions.
Furthermore we impose discrete symmetry 
under $x_1 \rightarrow x_2$, $x_2 \rightarrow x_1$, $x_4 \rightarrow -x_4$.
Then the parameters are restricted to 
\beq
t_1=t_2, \quad t_3=t_4 (\equiv \tilde{t}), \qquad u_1=u_2, \quad u_3=u_4=0 \ . 
\eeq


\section{Results}
\label{results}
For each solution of the self-consistency equations
obtained within the symmetry Ansatz,
we calculate the free energy. (See appendix for the details.)
The free energy we plot in what follows is actually defined by
\begin{equation}
  f = \lim_{N\to\infty} \left\{ F - 2(1-r)\ln N\right\} \ ,
\label{fe_density}
\end{equation}
where the subtraction is necessary to make the quantity finite.
The exact result at infinitesimal $r$ is given by \cite{exact}
\beq
f = - 2 \ln 2 + (1 - \ln 2) r  + {\rm o}(r) \ .
\label{f_exact}
\eeq

\FIGURE{\epsfig{file=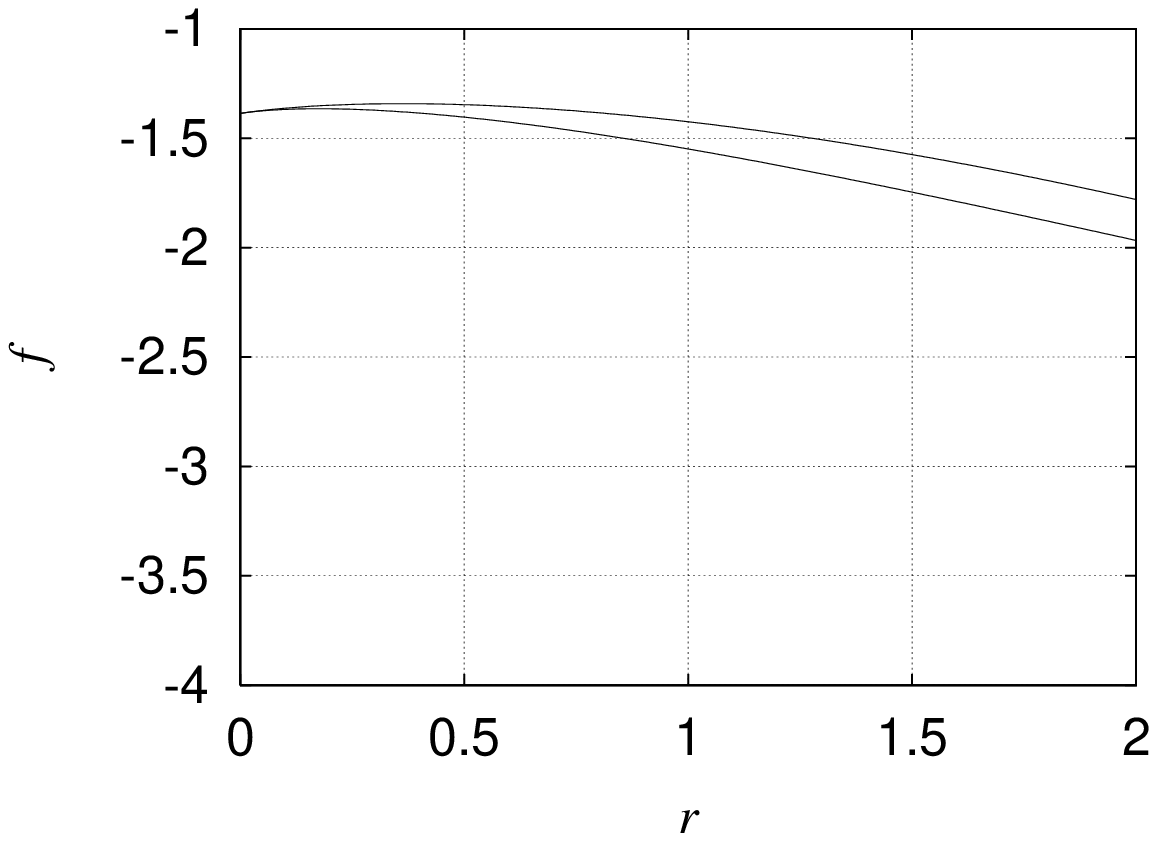,width=7.4cm}
\epsfig{file=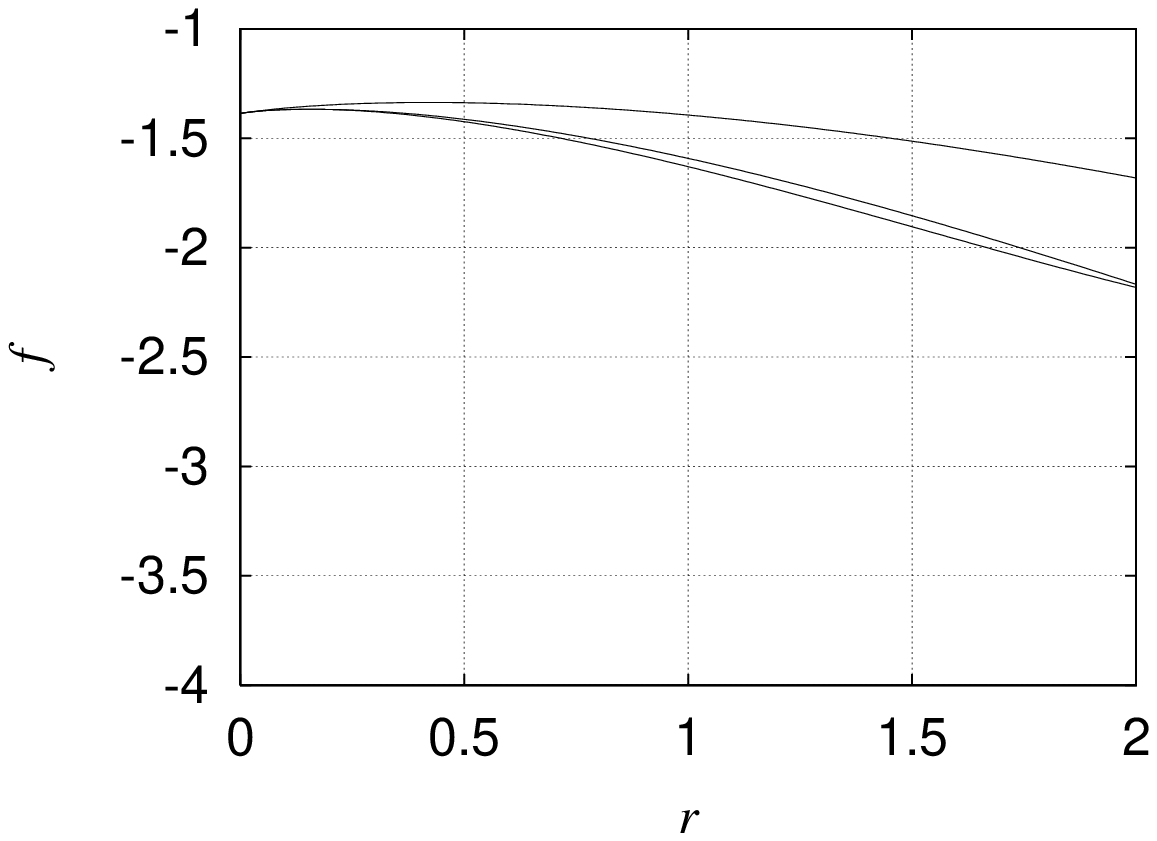,width=7.4cm}
\epsfig{file=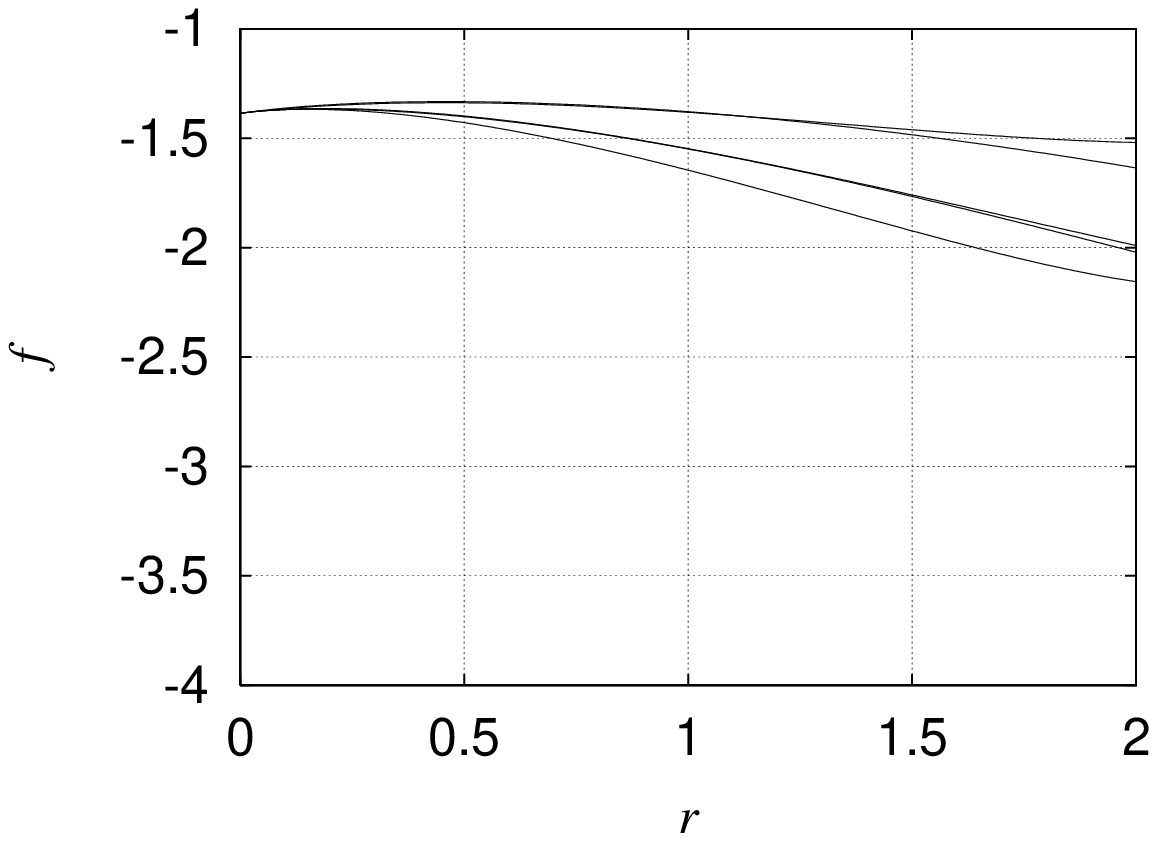,width=7.4cm}
\epsfig{file=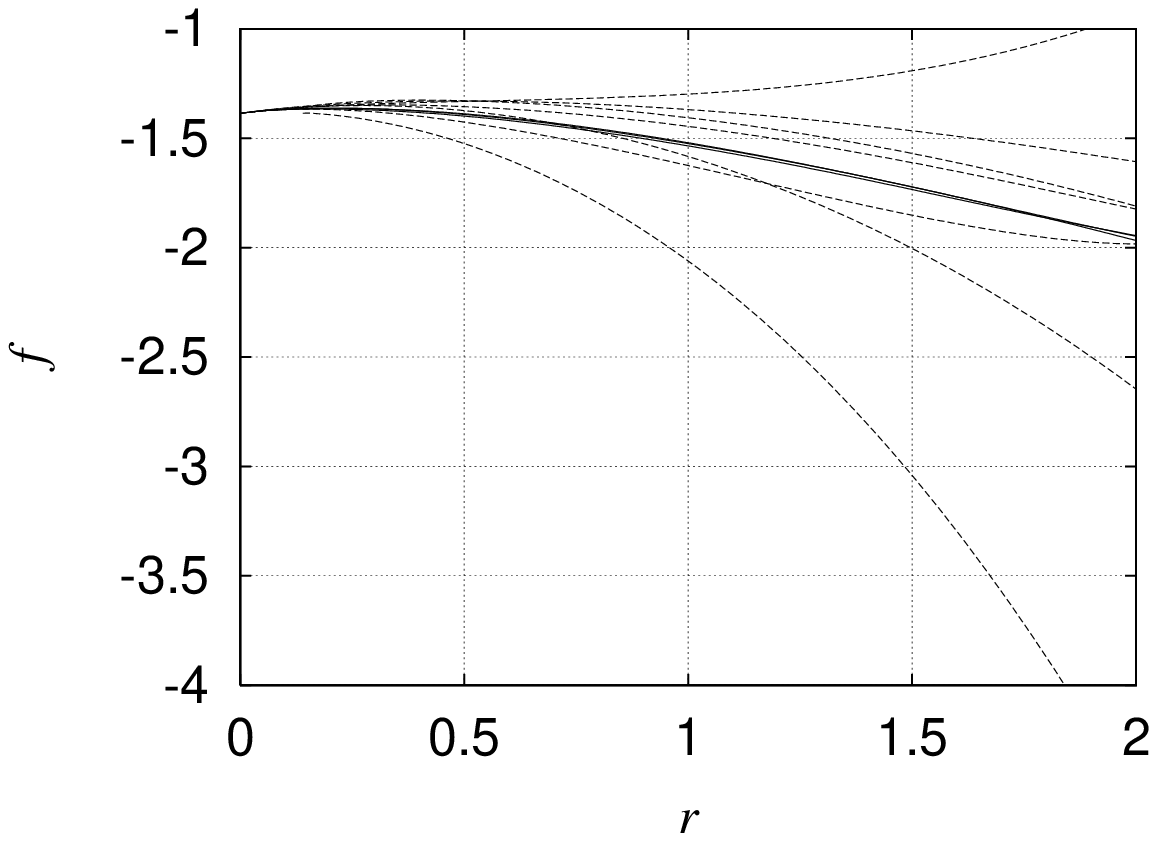,width=7.4cm}%
\caption{The free energy obtained for the SO(3) Ansatz 
at orders 3(left top), 5(right top), 7(left bottom) and 9(right bottom)
is plotted as a function of $r$.
\label{fig:freeSO3}}}

\FIGURE{\epsfig{file=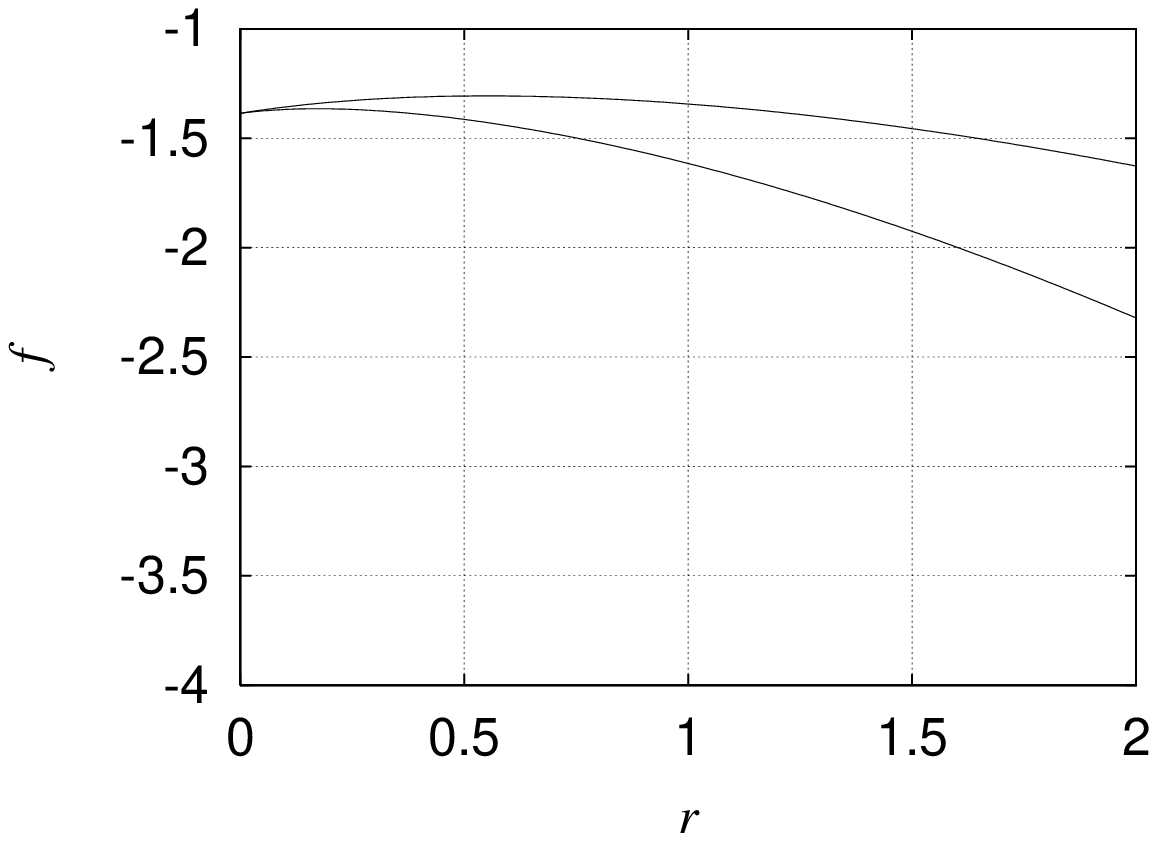,width=7.4cm}
\epsfig{file=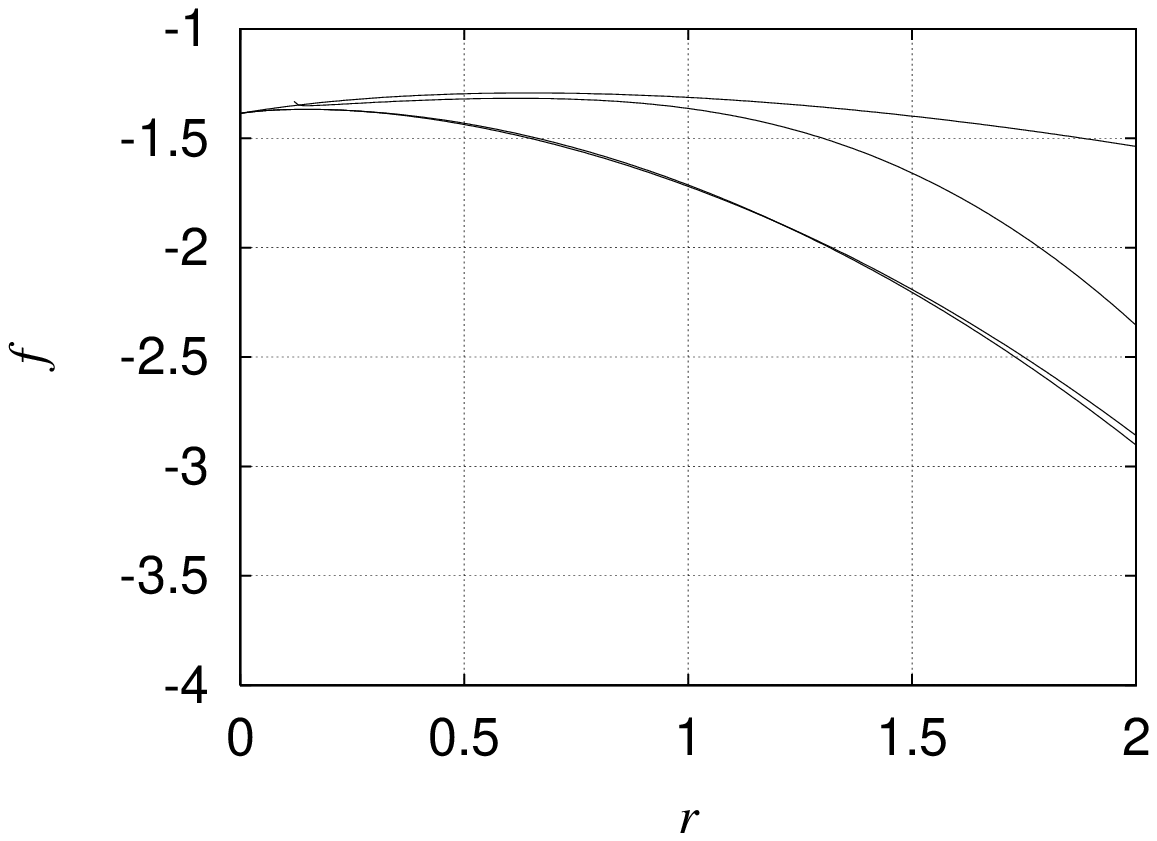,width=7.4cm}
\epsfig{file=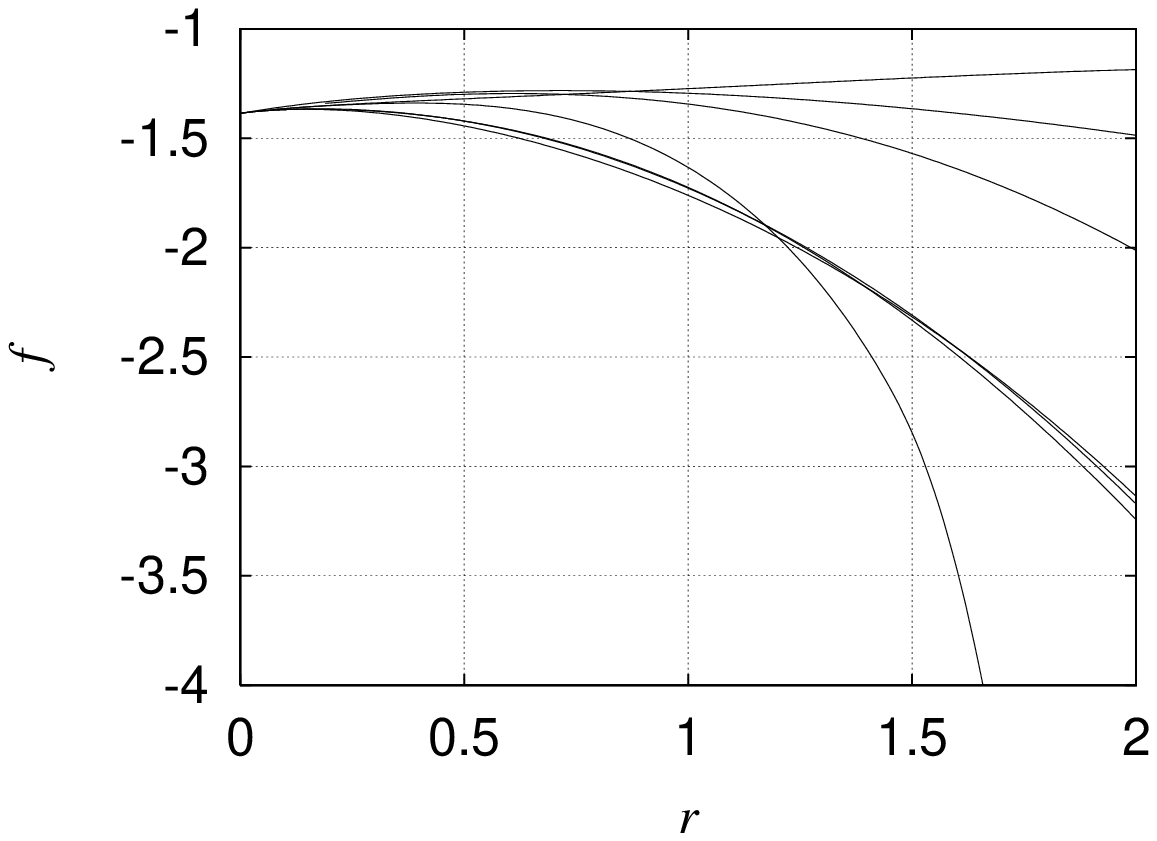,width=7.4cm}
\epsfig{file=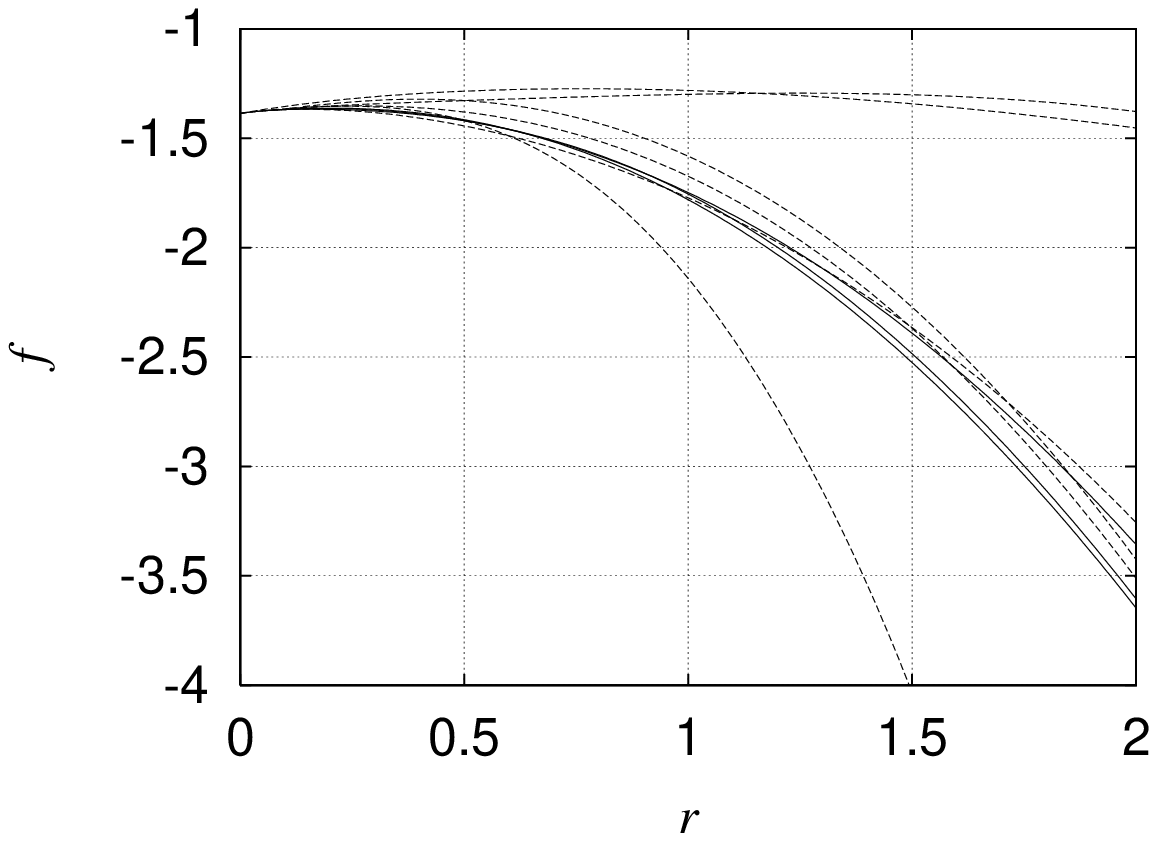,width=7.4cm}%
\caption{The free energy obtained for the SO(2) Ansatz
at orders 3(left top), 5(right top), 7(left bottom) and 9(right bottom)
is plotted as a function of $r$.
\label{fig:freeSO2}}}

In Fig.\ \ref{fig:freeSO3}
the free energy calculated for the SO(3) Ansatz at orders 3,5,7,9
is plotted against $r$. 
We find that at orders 5 and 7 there are two solutions which almost 
coincide with each other throughout the whole region 
\footnote{At $r \gtrsim 2$ the solutions that concentrate at the 9-th order
start to separate, and we cannot obtain reliable results in that regime. 
Note in this regard that
the number of fermion loops gives the power of
$r$ in the result of the gaussian expansion.
Therefore it is reasonable
that the convergence becomes slower as we go to larger $r$.}
of $0\le r \le 2$.
At the 9-th order there are actually three solutions lying on top of each other, 
which are represented by the
solid lines to be distinguished from the other solutions.
We consider this as an indication of the plateau formation.
Similar behavior is observed for the SO(2) Ansatz as one can see from
Fig.\ \ref{fig:freeSO2}. Again the three solid lines in the right bottom plot 
represent the solutions that we consider to be concentrating.

We pick up the three solutions which concentrate
at the 9-th order for the two Ans\"atze and 
plot them in Fig.\ \ref{fig:freeSO2SO3} for comparison.
Throughout the whole range of $r$ considered, the free energy for the
SO(2) Ansatz is smaller than that for the SO(3) Ansatz.
Thus we conclude that the true vacuum is described by the SO(2) Ansatz.
On the other hand, the results for the two Ans\"atze asymptote to each other
as $r$ approaches zero. The meaning of this behavior will be clarified shortly.

\FIGURE{\epsfig{file=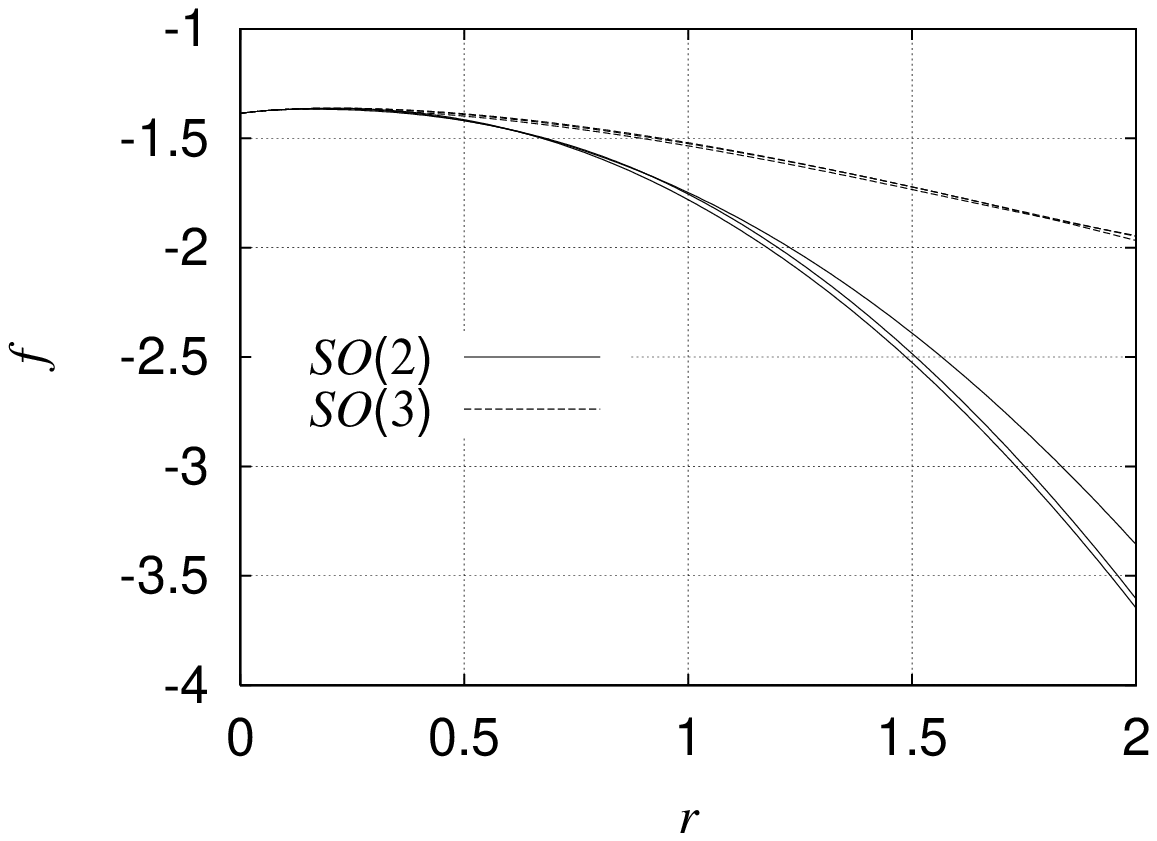,width=9cm}%
\caption{Comparison of the free energy obtained for the 
SO(2) and SO(3) Ans\"atze (solid lines and dashed lines, respectively).
The solutions that concentrate at the 9-th order are extracted from
Figs.\ \ref{fig:freeSO3} and \ref{fig:freeSO2}.
\label{fig:freeSO2SO3}}}

Let us move on to the calculation of observables.
Similarly to the free energy, we can calculate an observable
as an expansion with respect to $\lambda$ using the action
(\ref{GEM_action}).
In our model the observable of primary interest
is the eigenvalues $\lambda_i$ 
of the ``moment of inertia tensor'' (\ref{def-inertia}).
We calculate them for the SO(3) and SO(2) Ans\"atze at the 9-th order
as a function of the free parameters in the gaussian action,
and plug in the three solutions
that are seen to concentrate in the study of free energy.
In fact the VEVs of $\lambda_i$ can be readily obtained
by diagonalizing $c_{\mu\nu}$ defined by eq.\ (\ref{full_prop_A}),
which is calculated anyway in the calculation of free energy
\footnote{Since $c_{\mu\nu}$ takes the form 
(\ref{full_prop_SO3}) and (\ref{full_prop_SO2})
respectively for the SO(3) and SO(2) Ans\"atze,
the diagonalization is actually needed only for the SO(2) Ansatz.}.
Fig.\ \ref{fig:extentSO3SO2} shows the results.
Note that we are ultimately interested in the results
for the SO(2) Ansatz since it gives the smaller free energy.

For the SO(3) (SO(2)) Ansatz
the lines that grow almost linearly actually represent three (two)
eigenvalues, which are degenerate due to the assumed symmetry.
For the SO(2) Ansatz 
it turns out that the third largest eigenvalue comes
closer to the two degenerate largest ones
as one approaches $r=0$,
thus realizing the SO(3) symmetry asymptotically. 
In fact the results obtained for the SO(2) Ansatz are indistinguishable
from those obtained for the SO(3) Ansatz at small $r$.
This is consistent with our observation in 
Fig.\ \ref{fig:freeSO2SO3} that 
the free energy for the SO(2) Ansatz have the same asymptotic
behavior for $r \rightarrow 0$ as that for the SO(3) Ansatz.
Actually we find that both the free energy and the observable
agree asymptotically with the exact results 
(\ref{f_exact}), (\ref{SSBresult}) at infinitesimal $r$.

At large $r$, on the other hand, the results for the SO(2) Ansatz show 
a clear tendency that the two degenerate eigenvalues become much larger
than the other two, which implies the emergence of a two-dimensional 
``space-time''. This agrees with the argument
based on the phase stationarity given at the end of Section \ref{model}.
Thus the gaussian expansion method enables us to obtain results
at finite $r$, 
which naturally interpolate the SO(3) symmetric exact result at 
infinitesimal $r$ and the two-dimensional behavior expected at large $r$.

\FIGURE{\epsfig{file=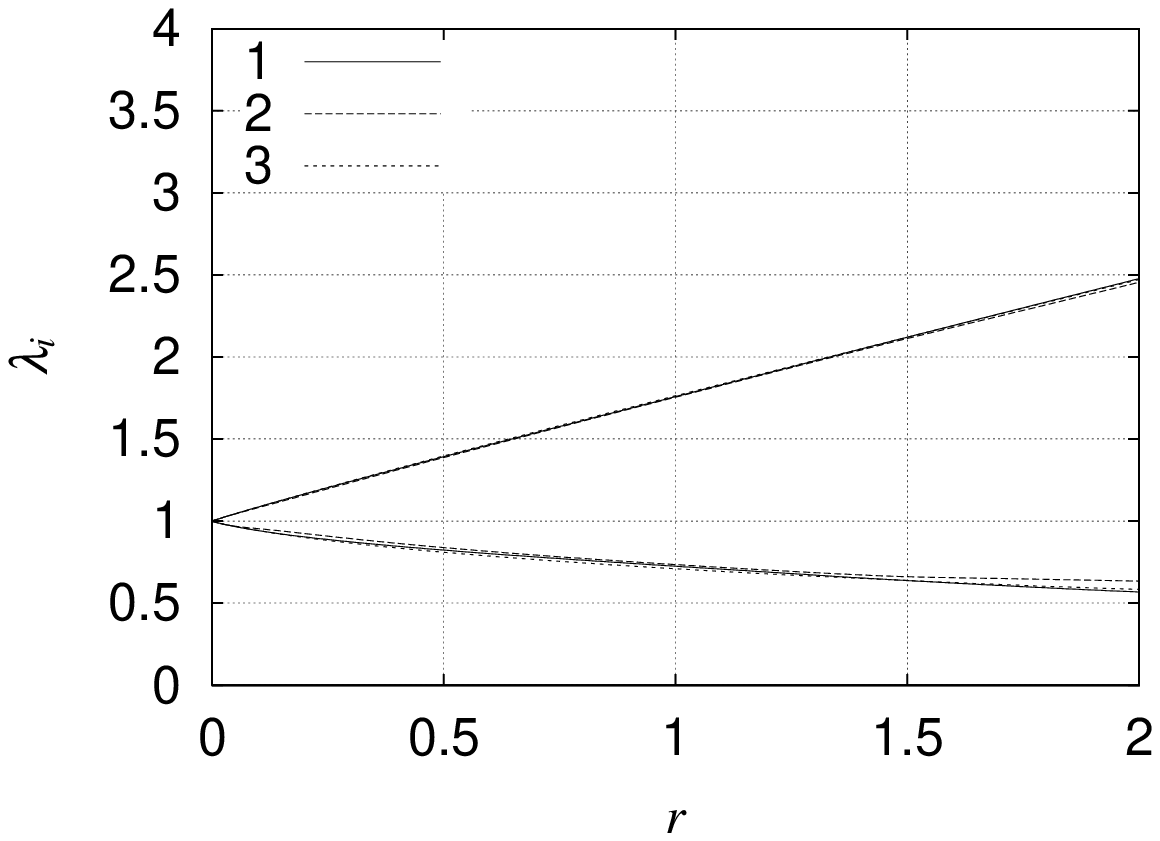,width=7.4cm}
\epsfig{file=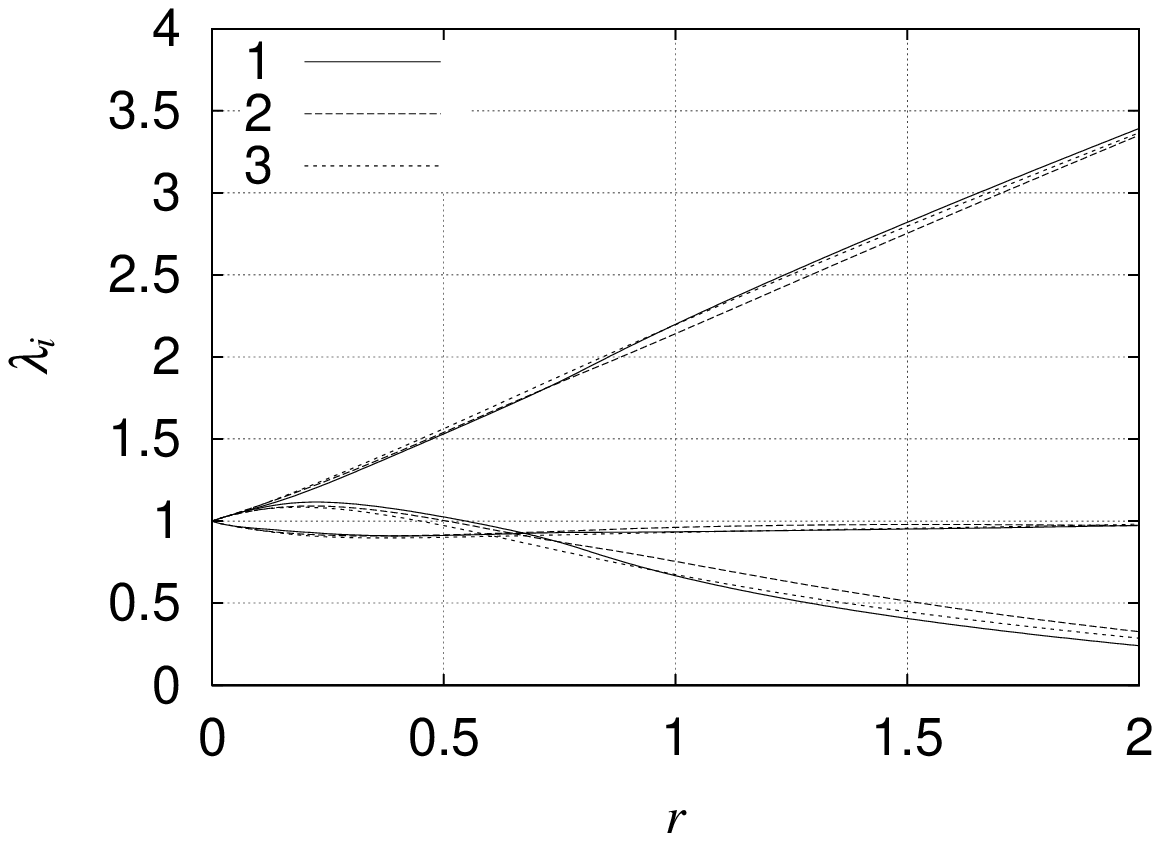,width=7.4cm}
\caption{The four eigenvalues of the ``moment of inertia tensor''
(\ref{def-inertia})
obtained for the SO(3) Ansatz (left) and the SO(2) Ansatz (right)
at the 9-th order are shown as a function of $r$.
Note that the largest eigenvalue
has 3-fold (2-fold) degeneracy for the SO(3) Ansatz and 
the SO(2) Ansatz, respectively.
The three types of line correspond to the three solutions of the 
self-consistency equations (\ref{self-consistency_eq})
that concentrate at the 9-th order. 
\label{fig:extentSO3SO2}}}



\section{Summary and Discussions}
\label{summary}

In this paper we have applied the gaussian expansion method
to a matrix model which is expected to exhibit SSB of
rotational symmetry due to the phase of the fermion determinant.
The free energy calculated by the gaussian expansion method
depends on the free parameters in the gaussian action, and 
the formation of a plateau in the parameter space is crucial
for the validity of the method.
For each Ansatz considered for the possible breaking pattern of
the SO(4) symmetry, we obtained
a clear evidence for the plateau formation.
By comparing the free energy obtained for each Ansatz,
we concluded that the true vacuum is described by the SO(2) Ansatz.
Our results for
the extent of ``space-time'' in each direction
are consistent with the exact result for
infinitesimal $r$ and also with the behavior expected at large $r$.

The mechanism for the SSB of rotational symmetry
demonstrated in the present model
is expected to be at work also in the IIB matrix model.
However, we should also note the difference of the two models.
In the present model the fermionic degrees of freedom match 
with the bosonic ones at $r=1$, but there the ``space-time'' is
not really two-dimensional, but it looks like a ``rugby ball''
with two directions more extended than the other two.
Moreover, if we consider the ten-dimensional version of the present model,
we expect that there will be eight directions more extended than the other 
two. In the IIB matrix model, on the other hand, the result of the 
gaussian expansion method
shows that the ratio of the extent of space-time in four directions
to that in the remaining six directions increases with the order
up to the 7-th order \cite{Nishimura:2001sx,KKKMS,KKKS}.
This suggests that {\em four} directions are {\em much more} 
(possibly, infinitely more) extended
than the remaining six directions. 
It is conceivable that supersymmetry plays an important role here.
Let us recall that the effective theory 
for the eigenvalues of the bosonic matrices in the IIB matrix model
is a weakly bound system like a branched polymer due to cancellation between the 
bosonic and fermionic contributions \cite{AIKKT}.
This makes the space-time easy to collapse.
Note also that the convergence of the gaussian expansion 
is not so clear in the IIB matrix model as in the present model.

While it is certainly worth while to proceed
to the 8-th or 9-th orders in the IIB matrix model, 
we consider that Monte Carlo simulations along the line of 
ref.\ \cite{sign} are necessary to definitely confirm the emergence of
a four-dimensional space time.
That approach is also expected to provide an intuitive understanding of
why ``4'' instead of 3 or 5.
Note in this regard that the present simplified model shares
the technical difficulty for implementing the phase of the fermion determinant
in Monte Carlo simulation.
Our new results obtained in this paper therefore provide
a nice testing ground for the new method for simulating the IIB matrix model.

\acknowledgments

The authors would like to thank K.N.\ Anagnostopoulos,
T.\ Aoyama and H.\ Kawai 
for fruitful discussions.

\bigskip

\appendix

\section{Details of the calculation}
\label{sec:SDmethod}
\setcounter{equation}{0}
\renewcommand{\theequation}{A.\arabic{equation}}
As explained in Section \ref{GEM} 
the main part of calculations in the gaussian expansion
is actually nothing but the ordinary perturbative calculation
of the free energy ${\cal F}(t,u)$ defined by (\ref{calF}).
The Feynman rules 
are given in 
Fig. \ref{fig:feynman}.
The first and second lines represent the bare propagators 
$\langle (A_{\mu})_{ij}(A_{\nu})_{kl}\rangle_0$ and
$\langle \psi_{\alpha}^{fi}\bar{\psi}_{\beta}^{gj}\rangle_0$, respectively,
where the symbol $\langle  \ \cdot \ \rangle_0$
represents a VEV obtained with the gaussian action (\ref{defS0}).
The third line stands for the interaction vertex coming from $S_{{\rm f}}$. 
Instead of evaluating all the diagrams contributing to 
${\cal F}(t,u)$, we use the SD equations 
to reduce the number of the diagrams 
to be computed following the idea put forward in ref.~\cite{KKKMS}. 

\FIGURE{\epsfig{file=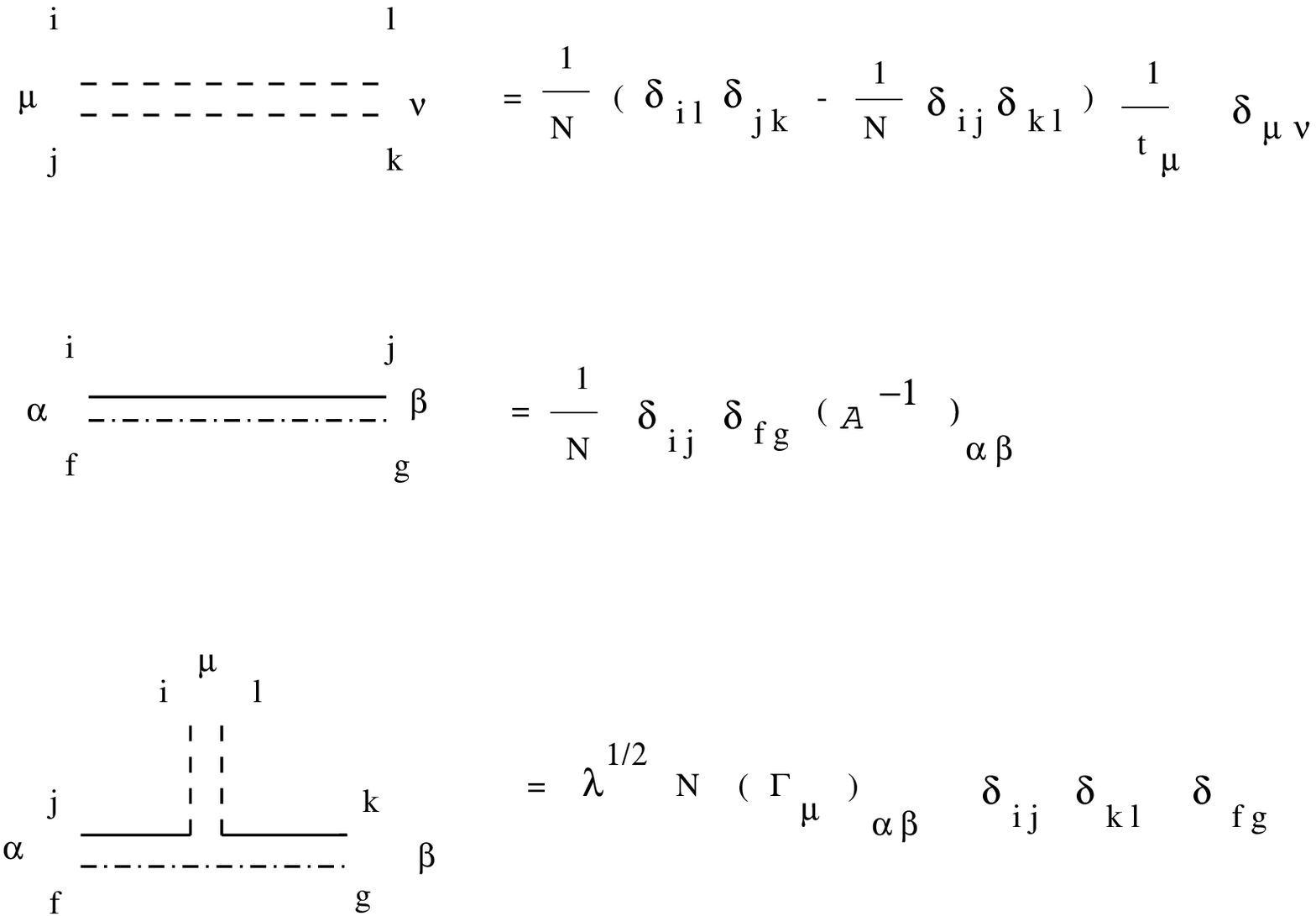,height=7cm}
    \caption{The Feynman rules for the $\lambda$-expansion 
of (\ref{calF}).}
    \label{fig:feynman}
}

Note that the one-point function $\langle (A_{\mu})_{ij}\rangle$ is proportional to 
$ \delta_{ij}\langle \tr A_{\mu}\rangle$ because of the SU($N$) symmetry and 
therefore it vanishes due to the tracelessness condition imposed on $A_{\mu}$.
The full propagators (connected two-point functions) can be written in the form 
\beqa
\label{full_prop_A}
\langle (A_{\mu})_{ij}(A_{\nu})_{kl}\rangle & = & 
\frac{1}{N}
\left(\delta_{il}\delta_{jk}-\frac{1}{N}\delta_{ij}\delta_{kl}\right)
c_{\mu\nu} \ ,  \\
\langle \psi_{\alpha}^{fi}\bar{\psi}_{\beta}^{gj}\rangle & = & 
\frac{1}{N}\delta_{fg}\delta_{ij}{\cal D}_{\alpha\beta} \ ,
\eeqa
where the coefficients $c_{\mu\nu}$ and ${\cal D}$ are given
at the leading order in $\lambda$ as
\beq
c_{\mu\nu} = \frac{1}{t_{\mu}}\delta_{\mu\nu} + \mbox{O}(\lambda) \ , \quad 
{\cal D} = {\cal A}^{-1} + \mbox{O}(\lambda) \ . 
\eeq
Corresponding to eq.\ (\ref{defA}),
the full propagator ${\cal D}$ can be parametrized as
\beq
{\cal D}= \sum_{\mu =1}^4 d_{\mu}\bar{\Gamma}_{\mu} \ ,
\eeq
where the ``conjugate'' gamma matrices $\bar{\Gamma}_{\mu}$ 
are defined by 
\beq
\bar{\Gamma}_k = \Gamma_k \quad (k=1,2,3) \ , \qquad \bar{\Gamma}_4=-\Gamma_4 \ .
\eeq
In what follows, we restrict ourselves to the large-$N$ limit 
with the ratio $r=\frac{N_{\rm f}}{N}$ fixed, 
so that we have to consider planar diagrams only, 
but the method itself is applicable to finite $N$ 
as well.

By using the SD equations for the full propagators, we can 
reduce the calculation of the free energy to that of 
two-particle-irreducible (2PI) planar diagrams. 
Here, by ``2PI diagrams'' we mean those diagrams which cannot be 
separated into two disconnected parts 
by cutting two propagators. Let us consider the 2PI planar vacuum diagrams 
whose internal lines are all replaced by the full propagators. 
The sum of such diagrams is a function of 
$c_{\mu\nu}$ and $d_\mu$ and shall be denoted as $N^2 G(c, d)$. 
For example, $N^2G(c, d)$ up to the 5th order
can be computed from the 5 diagrams in Fig. \ref{fig:2PI}, 
and it is given explicitly as
\beqa
G(c, d) & = & 
- \, \frac12 \, \lambda  \,  r 
\sum_{\mu\nu}
c_{\mu\nu}\, \Tr\Bigl(\Gamma_{\mu}{\cal D}\Gamma_{\nu}{\cal D}\Bigr) 
\nonumber \\
& & + \, \frac16 \, \lambda^3  \,  r^2
\sum_{\mu \nu \lambda \rho \kappa \sigma}
c_{\mu\nu}c_{\lambda\rho}c_{\kappa\sigma} \, 
\Tr\Bigl(\Gamma_{\mu}{\cal D}\Gamma_{\lambda}{\cal D}\Gamma_{\kappa}{\cal D}\Bigr)
 \, 
\Tr\Bigl(\Gamma_{\sigma}{\cal D}\Gamma_{\rho}{\cal D}
\Gamma_{\nu}{\cal D}\Bigr) \nonumber \\
& & + \,  \frac18 \, \lambda^4 \, r^2 
\sum_{\mu \nu \lambda \rho \kappa \sigma  \tau  \xi} 
c_{\mu\nu}c_{\lambda\rho}c_{\kappa\sigma}c_{\tau\xi} \, 
\Tr\Bigl(\Gamma_{\mu}{\cal D}\Gamma_{\lambda}{\cal D}
\Gamma_{\kappa}{\cal D}\Gamma_{\tau}{\cal D}\Bigr)
\, \Tr\Bigl(\Gamma_{\xi}{\cal D}\Gamma_{\sigma}{\cal D}
\Gamma_{\rho}{\cal D}\Gamma_{\nu}{\cal D}\Bigr)
\nonumber \\
& &  + \,\frac{1}{10} \, \lambda^5 \,  r^2 
\sum_{\mu \nu \lambda \rho \kappa \sigma  \tau  \xi  \eta  \zeta} 
c_{\mu\nu}c_{\lambda\rho}c_{\kappa\sigma}c_{\tau\xi}c_{\eta\zeta} \, 
\Tr\Bigl(\Gamma_{\mu}{\cal D}\Gamma_{\lambda}{\cal D}
\Gamma_{\kappa}{\cal D}\Gamma_{\tau}{\cal D}
\Gamma_{\eta}{\cal D}\Bigr)
\nonumber \\
& &  \hspace{6.5cm}\times
\Tr\Bigl(\Gamma_{\zeta}{\cal D}\Gamma_{\xi}{\cal D}
\Gamma_{\sigma}{\cal D}\Gamma_{\rho}{\cal D}
\Gamma_{\nu}{\cal D}\Bigr)
\nonumber \\
& & 
- \,  \frac12 \, \lambda^5 \, r^3 
\sum_{\mu \nu \lambda \rho \kappa \sigma \tau  \xi \eta \zeta} 
c_{\mu\nu}c_{\lambda\rho}c_{\kappa\sigma}c_{\tau\xi}c_{\eta\zeta} \, 
\Tr\Bigl(\Gamma_{\mu}{\cal D}\Gamma_{\lambda}{\cal D}
\Gamma_{\kappa}{\cal D}\Gamma_{\tau}{\cal D}\Bigr)
\, \Tr\Bigl(\Gamma_{\nu}{\cal D}\Gamma_{\rho}{\cal D}\Gamma_{\eta}{\cal D}\Bigr)
\nonumber \\
& & \hspace{6.5cm} \times
\Tr\Bigl(\Gamma_{\sigma}{\cal D}\Gamma_{\xi}{\cal D}\Gamma_{\zeta}{\cal D}\Bigr) 
\nonumber \\
& & + \,  \mbox{O}(\lambda^6) \ , 
\eeqa
where ``$\Tr$'' implies a trace taken with respect to the spinor indices.

\FIGURE{\epsfig{file=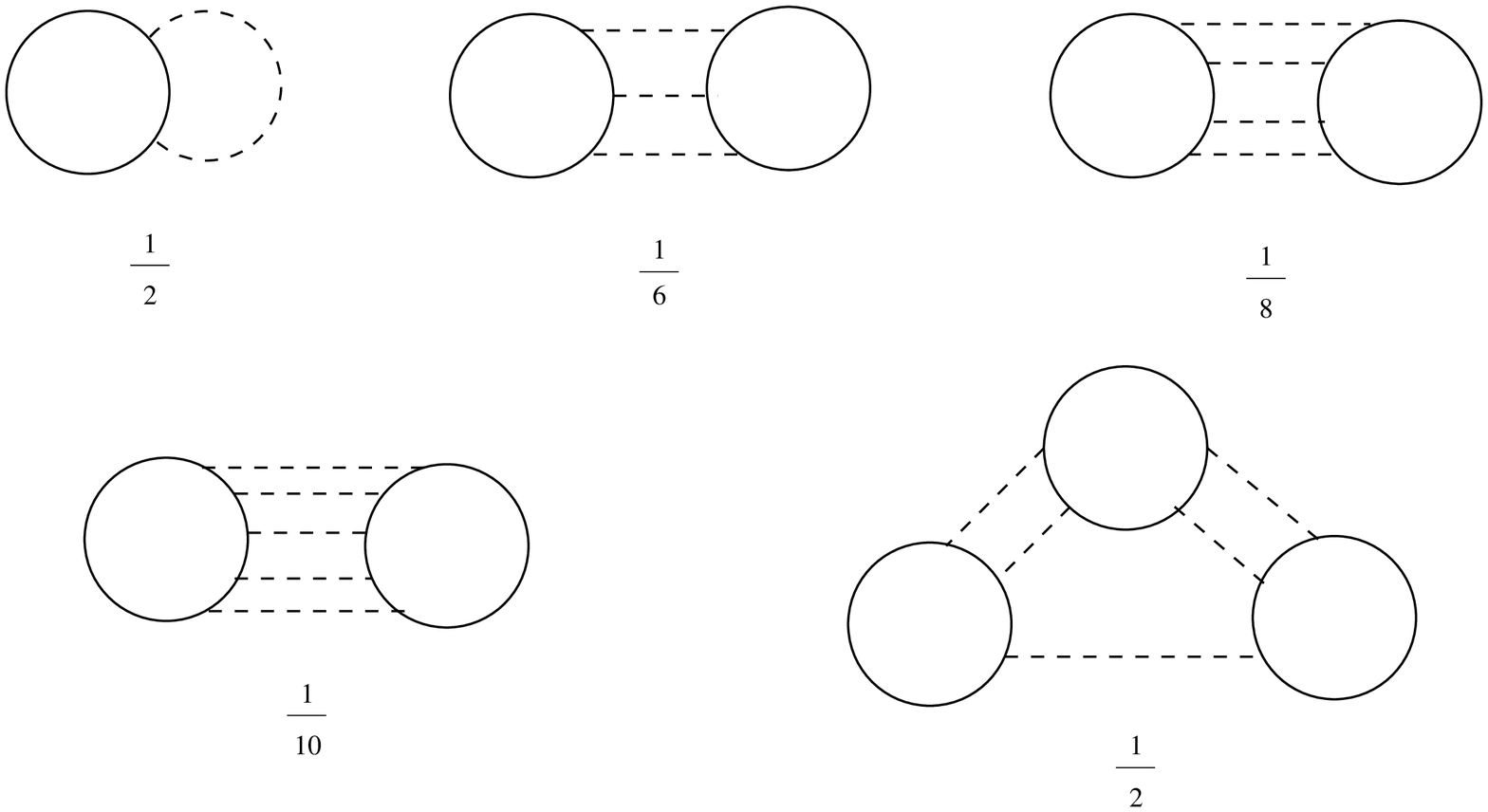,height=6.5cm}
    \caption{The 2PI planar vacuum diagrams
up to the 5-th order.
For simplicity the propagators for bosons and fermions are represented
by single (dashed and solid, respectively) lines. 
The first three diagrams are contributions from the orders 1, 3, 4, and 
the last two from the order 5. (There is no contribution from the 2nd order.) 
The symmetry factor is indicated below each diagram. 
}
    \label{fig:2PI}
}

The complete list of 2PI planar vacuum diagrams at orders 6, 7, 8, 9
are given in Figs.\ \ref{fig:2PI_6}-\ref{fig:2PI_9-4}.
The number of diagrams at each order is 4, 9, 24, 81, respectively.
   
\FIGURE{\epsfig{file=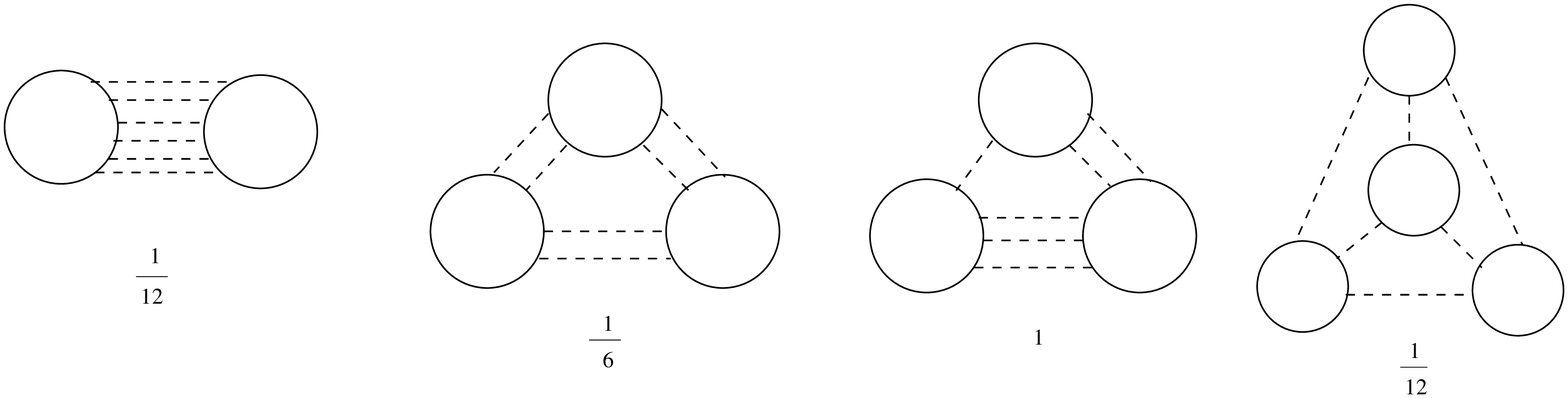,height=3.5cm}
    \caption{The 2PI planar vacuum diagrams at the 6-th order.
}
    \label{fig:2PI_6}
}
 
\FIGURE{\epsfig{file=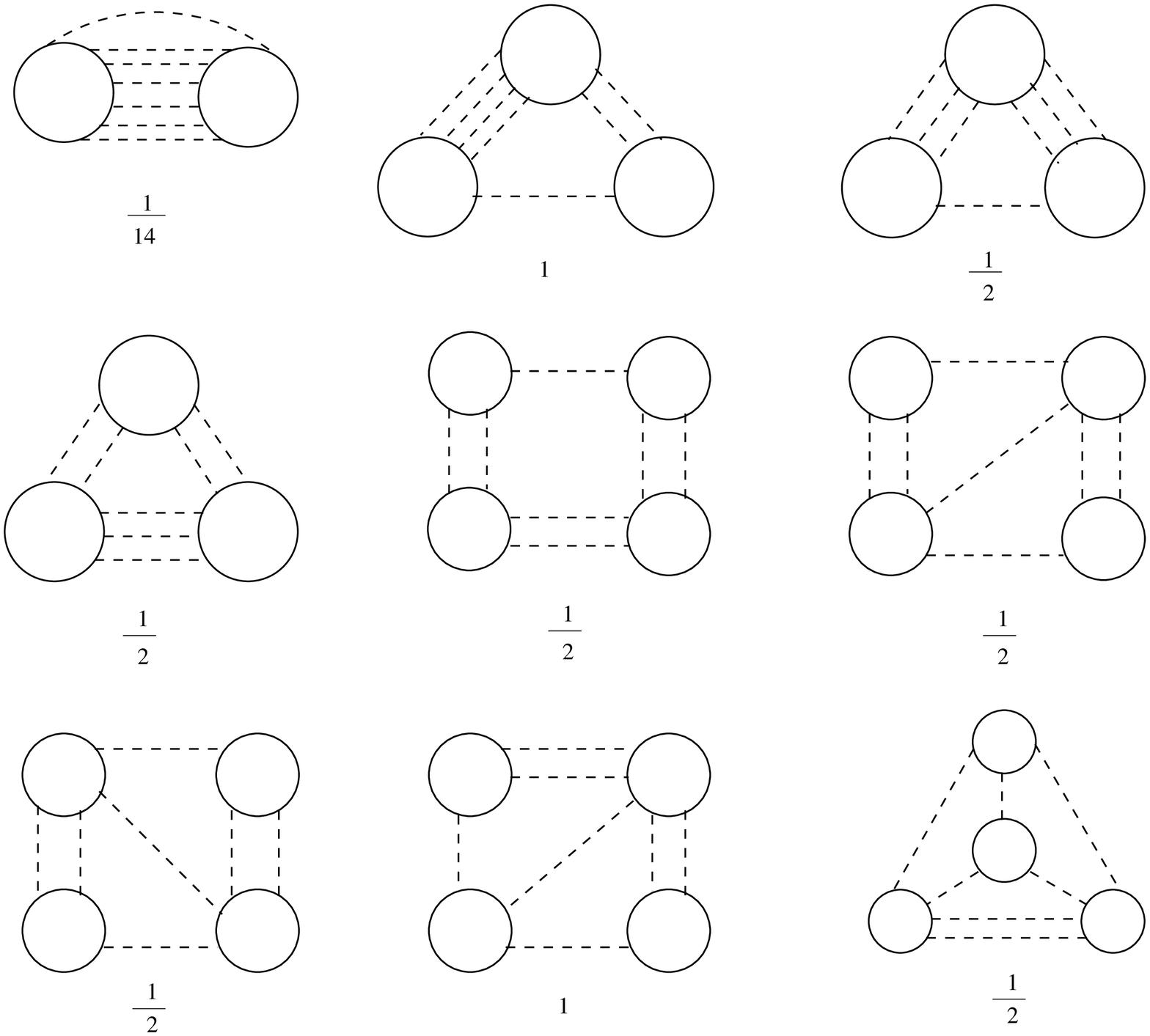,height=10cm}
    \caption{The 2PI planar vacuum diagrams at the 7-th order.
}
    \label{fig:2PI_7}
}

\FIGURE{\epsfig{file=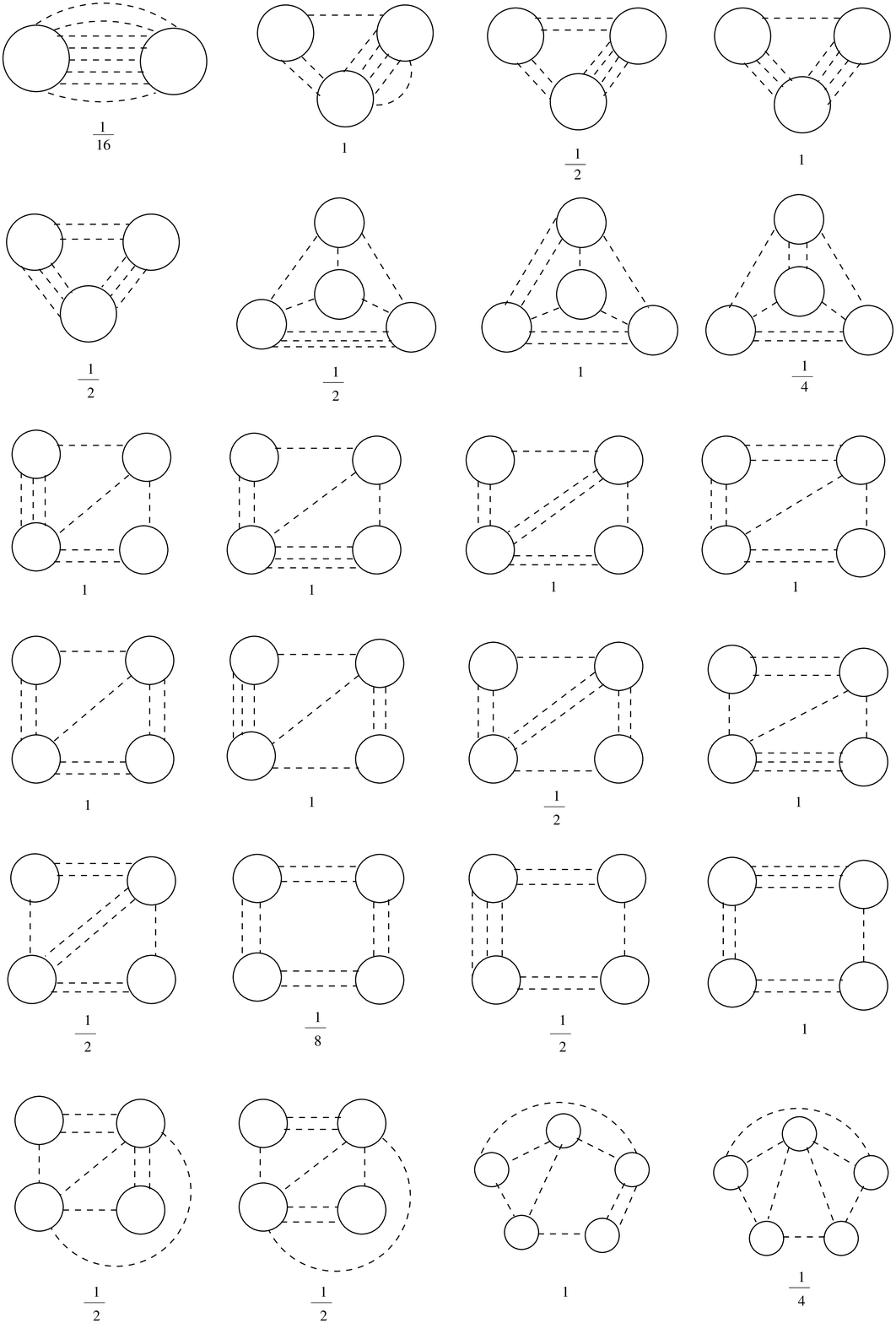,height=20cm}
    \caption{The 2PI planar vacuum diagrams at the 8-th order.
}
    \label{fig:2PI_8}
}

\FIGURE{\epsfig{file=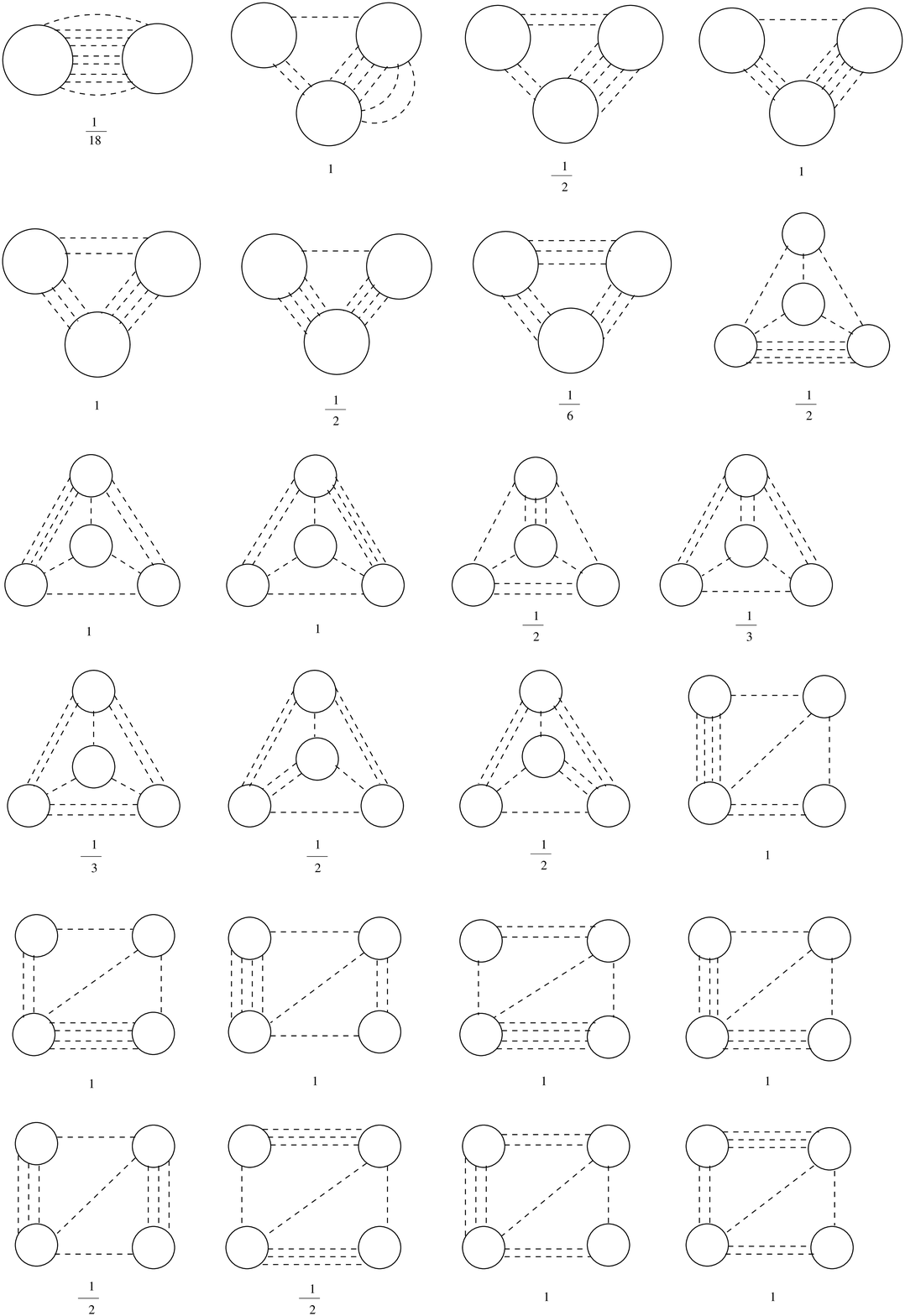,height=20cm}
    \caption{The 2PI planar vacuum diagrams at the 9-th order (to be continued). 
}
    \label{fig:2PI_9-1}
}

\FIGURE{\epsfig{file=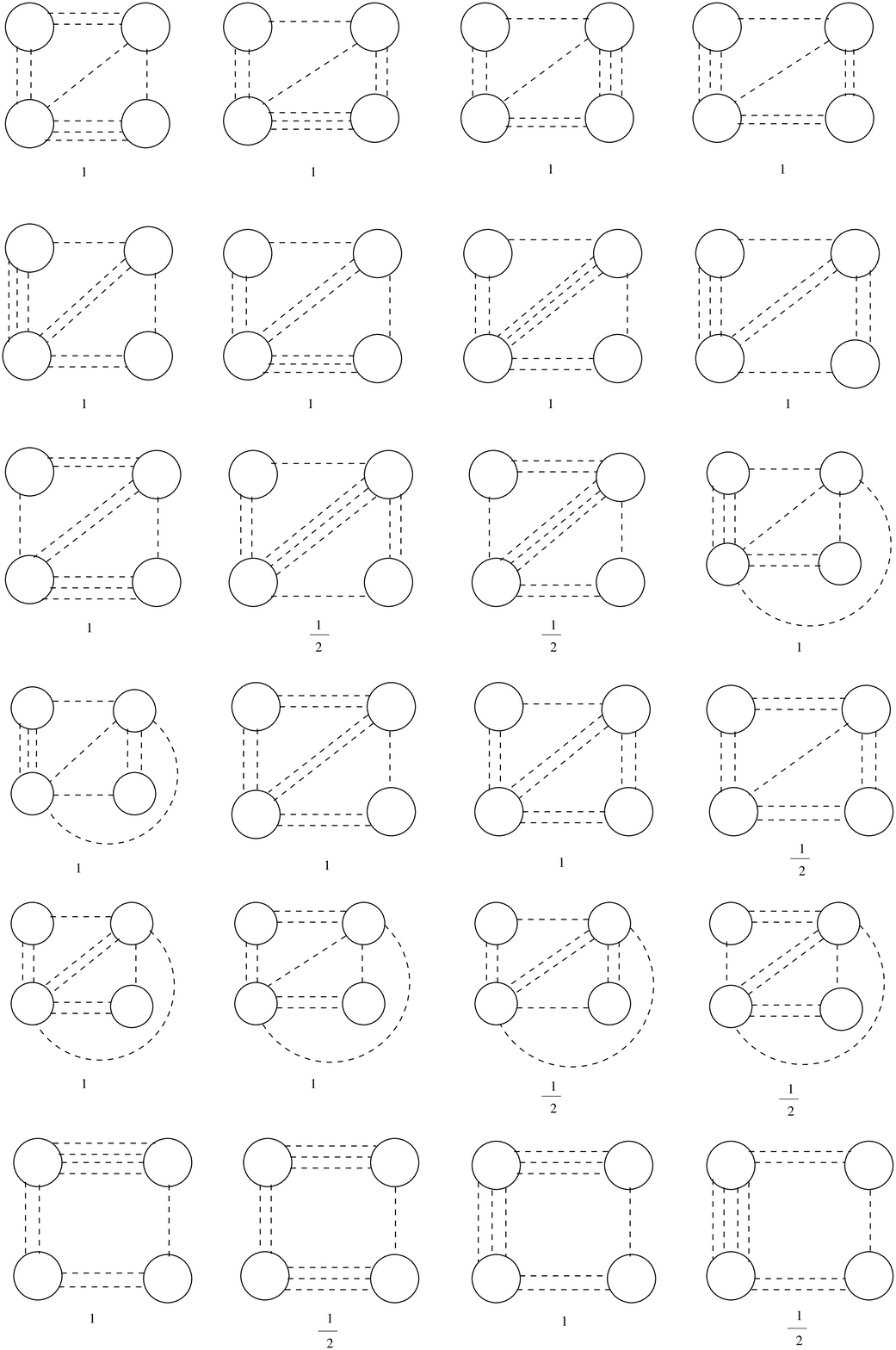,height=20cm}
    \caption{The 2PI planar vacuum diagrams at the 9-th order (to be continued). 
}
    \label{fig:2PI_9-2}
}

\FIGURE{\epsfig{file=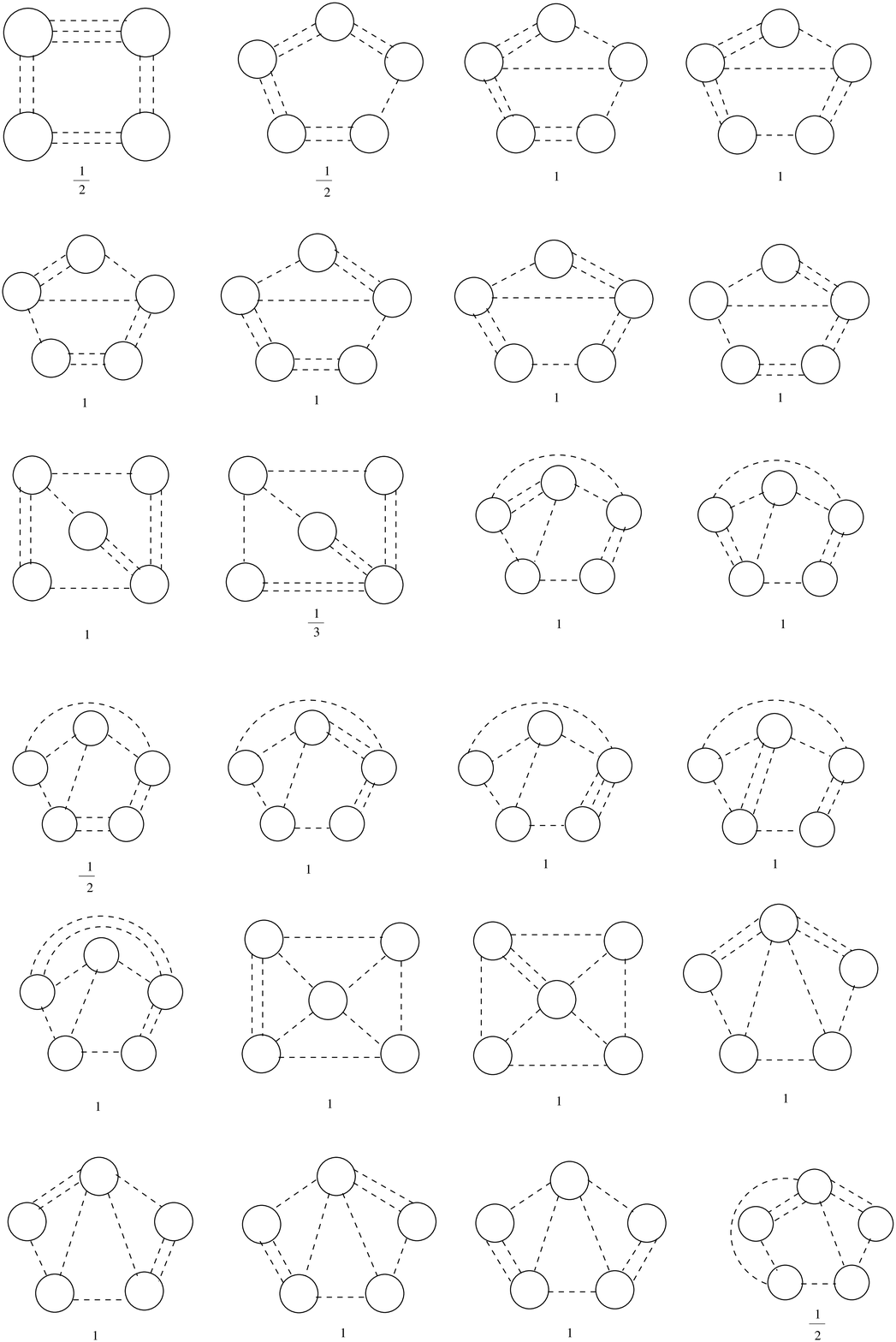,height=20cm}
    \caption{The 2PI planar vacuum diagrams at the 9-th order (to be continued). 
}
    \label{fig:2PI_9-3}
}

\FIGURE{\epsfig{file=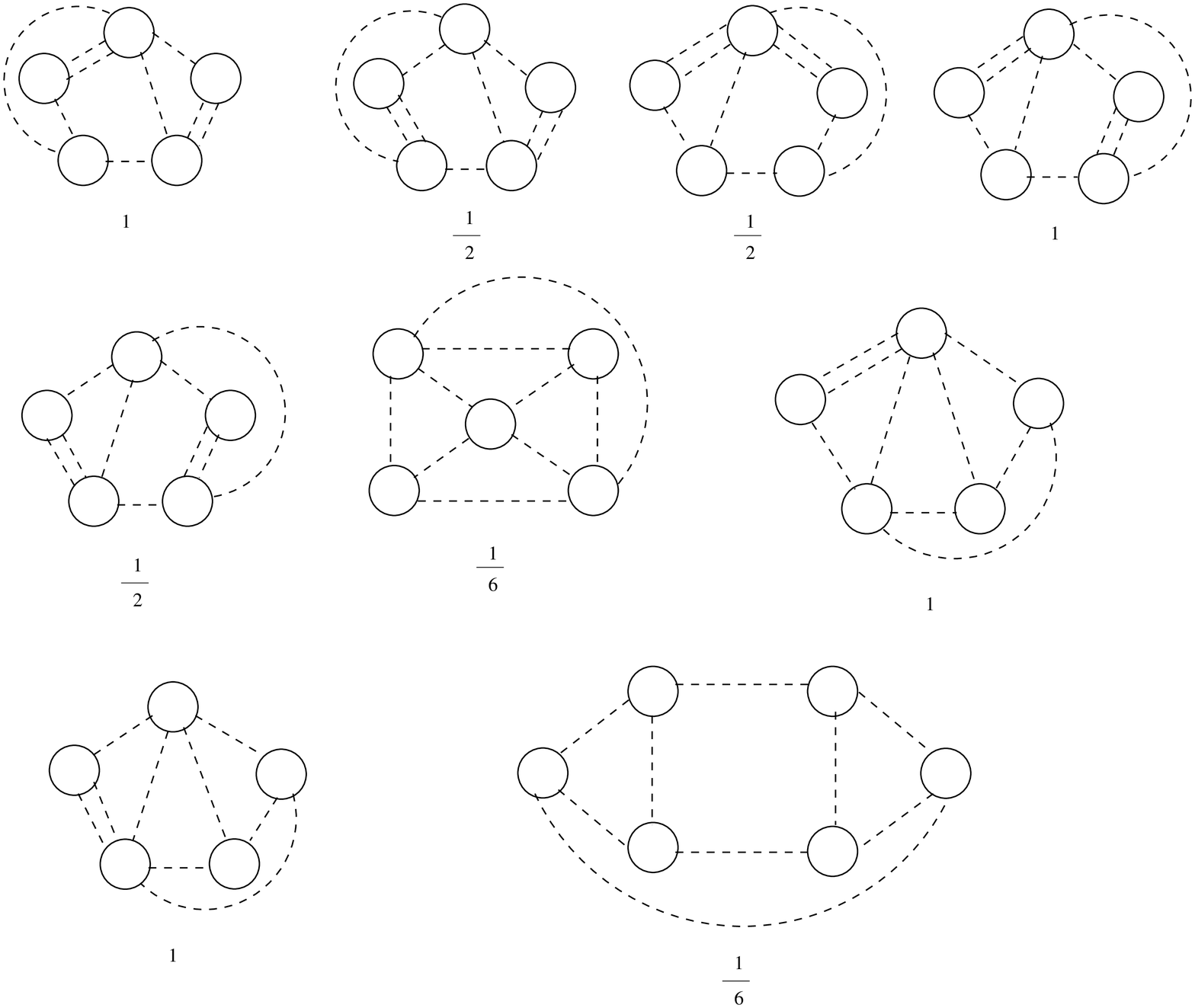,height=10cm}
    \caption{The 2PI planar vacuum diagrams at the 9-th order.
}
    \label{fig:2PI_9-4}
}

\subsection{Schwinger-Dyson equations}
\label{sec:SDeq}
In this section we derive a closed set of equations, which 
allows us to calculate ${\cal F}(t,u)$ from $G(c,d)$.

Let $B_{\mu\nu}$ be the one-particle-irreducible (1PI) part 
of the radiative corrections to $c_{\mu\nu}$. 
Then we can write $c_{\mu\nu}$ as a geometric series 
\beq
c = \frac{1}{T} + \frac{1}{T} \, B \, \frac{1}{T} + 
\frac{1}{T} \, B \, \frac{1}{T} \, B \, \frac{1}{T} + \cdots 
  = \frac {1}{T -B} \ , 
  \label{SDc1}
\eeq
where we regard 
$c_{\mu\nu}$, $T_{\mu\nu}\equiv t_{\mu}\delta_{\mu\nu}$, $B_{\mu\nu}$ 
as $4\times 4$ matrices.
Note, on the other hand, 
that $B_{\mu\nu}$ can be obtained 
from $G(c, d)$ by 
\beq
B_{\mu\nu} = 2 \, \frac{\del}{\del c_{\mu\nu}} G(c, d) \ .  
\label{SDc2}
\eeq 
Combining (\ref{SDc1}) and (\ref{SDc2}), we obtain
\beq
(c^{-1})_{\mu\nu} = 
t_{\mu}\delta_{\mu\nu} - 2\frac{\del}{\del c_{\mu\nu}} G(c, d) \ . 
\label{SDc}
\eeq
Similarly we get 
\beq
({\cal D}^{-1})_{\alpha\beta} = {\cal A}_{\alpha\beta} 
+ \frac{1}{r}\frac{\del}{\del {\cal D}_{\beta\alpha}} G(c, d) \ ,
\label{SDD}
\eeq
which may be written in terms of $d_{\mu}$ as
\beq
 \frac{1}{\Delta}d_{\mu} = 
u_{\mu} + \frac{1}{2r}\frac{\del}{\del d_{\mu}}G(c, d) \ , 
\label{SDd} 
\eeq 
where $\Delta=\sum_{\mu} (d_{\mu})^2$.

Since the SD equations (\ref{SDc}), (\ref{SDD}) are 
closed with respect to $c_{\mu\nu}$, ${\cal D}_{\alpha\beta}$, 
we can solve them order by order in $\lambda$ 
once the sum of the 2PI diagrams $G(c, d)$ is obtained. 
For example, the results for the first few orders are as follows. 
\beqa
c_{\mu\nu} & = & \sum_{n=0}^{\infty} \lambda^n \, 
c_{\mu\nu}^{(n)} \ , \\
c_{\mu\nu}^{(0)} & = & \frac{1}{t_{\mu}} \, 
 \delta_{\mu\nu} \ , \nonumber \\
c_{\mu\nu}^{(1)} & = & - r \, \frac{1}{t_{\mu}} \, \frac{1}{t_{\nu}}
 \Tr\Bigl(\Gamma_{\mu}{\cal A}^{-1}\Gamma_{\nu}{\cal A}^{-1}\Bigr)
\ , \nonumber \\
c_{\mu\nu}^{(2)} & = & - r
\sum_{\rho} 
\frac{1}{t_{\mu}} \, 
 \frac{1}{t_{\rho}} \, \frac{1}{t_{\nu}} \, 
 \Tr\left[\left\{\Gamma_{\mu}{\cal A}^{-1}, \Gamma_{\nu}{\cal A}^{-1}\right\} 
 \left(\Gamma_{\rho}{\cal A}^{-1}\right)^2\right] \nonumber \\
  & & + r^2
\sum_{\rho}
\frac{1}{t_{\mu}} \, \frac{1}{t_{\rho}} \, \frac{1}{t_{\nu}} \, 
 \Tr\Bigl(\Gamma_{\mu}{\cal A}^{-1}\Gamma_{\rho}{\cal A}^{-1}\Bigr)  
 \Tr\Bigl(\Gamma_{\rho}{\cal A}^{-1}\Gamma_{\nu}{\cal A}^{-1}\Bigr) \ , \nonumber \\
{\cal D}_{\alpha\beta} & = & \sum_{n=0}^{\infty} 
\lambda^n {\cal D}_{\alpha\beta}^{(n)} \ , \\
{\cal D}_{\alpha\beta}^{(0)} & = & 
\left({\cal A}^{-1}\right)_{\alpha\beta} \ , \nonumber \\
{\cal D}_{\alpha\beta}^{(1)} & = & 
\sum_{\mu}
\frac{1}{t_{\mu}}
\left[{\cal A}^{-1}\left(\Gamma_{\mu}{\cal A}^{-1}\right)^2\right]_{\alpha\beta} \ , 
 \nonumber \\
{\cal D}_{\alpha\beta}^{(2)} & = & 
\sum_{\mu\nu}
\frac{1}{t_{\mu}} \, 
\frac{1}{t_{\nu}}
\left[{\cal A}^{-1}\Gamma_{\mu}{\cal A}^{-1}\left(\Gamma_{\nu}{\cal A}^{-1}\right)^2
\Gamma_{\mu}{\cal A}^{-1}\right]_{\alpha\beta} 
\nonumber \\
 & & + 
\sum_{\mu\nu}\frac{1}{t_{\mu}} \, \frac{1}{t_{\nu}}
\left[{\cal A}^{-1}\left(\Gamma_{\mu}{\cal A}^{-1}\right)^2 
\left(\Gamma_{\nu}{\cal A}^{-1}\right)^2\right]_{\alpha\beta} \ . \nonumber
\eeqa  



Let us note that the free energy ${\cal F}(t,u)$ 
satisfies the differential equations 
\beqa
2\frac{\del}{\del t_{\mu}}{\cal F}(t,u) & = &   c_{\mu\mu} \ , \label{diffc} \\
 -\frac{1}{2r}\frac{\del}{\del u_{\mu}}{\cal F}(t,u) &=&  d_{\mu} \ .
\label{diffd}
\eeqa 
By integrating 
these equations term by term,
we finally obtain the $\lambda$-expansion for ${\cal F}(t,u)$.
The integration constant, which appears only in the zeroth order contributions 
(i.e., the one-loop diagrams), 
can be fixed easily by direct computation. 

The explicit form of the free energy up to the 2nd order is
\beqa
{\cal F}(t,u) & = & \sum_{k=0}^{\infty} \lambda^k {\widetilde {\cal F}}_k \ , \\
 {\widetilde {\cal F}}_0 & = & 
  2(1-r)\ln N -2\ln 2 + \frac12 \sum_{\mu}\ln t_{\mu} 
 -r \ln (\det {\cal A}) \ , 
\nonumber \\
{\widetilde {\cal F}}_1 & = & \frac{r}{2}\sum_{\mu}\frac{1}{t_{\mu}}
 \Tr\left[\left(\Gamma_{\mu}{\cal A}^{-1}\right)^2\right] \ , \nonumber \\
{\widetilde {\cal F} }_2 & = & 
\frac{r}{2}\sum_{\mu\nu}\frac{1}{t_{\mu}}\frac{1}{t_{\nu}}
 \Tr\left[\left(\Gamma_{\mu}{\cal A}^{-1}\right)^2
\left(\Gamma_{\nu}{\cal A}^{-1}\right)^2\right]
 -\frac{r^2}{4}\sum_{\mu\nu}\frac{1}{t_{\mu}}\frac{1}{t_{\nu}}
 \left(\Tr \Bigl(\Gamma_{\mu}{\cal A}^{-1}
\Gamma_{\nu}{\cal A}^{-1}\Bigr) \right)^2 \ . \nonumber
\eeqa



\subsection{Exploiting the Ans\"atze}
\label{sec:SD_ansatze}

As explained in Section \ref{GEM}
we impose some Ansatz on the possible symmetry breaking pattern
in order to reduce the number of free parameters in the gaussian action.
In this section we write down the SD equations
for each Ansatz separately.
By solving the SD equations, 
we obtain the free energy 
similarly to what we have done in section \ref{sec:SDeq}.

\subsubsection{SD equations for the SO(3) Ansatz}

The full propagators take the following form.
\beq
c_{\mu\nu}=\left(\begin{array}{cccc} 
c_{11} &           &           &         \\
       & \tilde{c} &           &         \\
       &           & \tilde{c} &         \\
       &           &           & \tilde{c} \end{array} 
\right) \ , \qquad d_2=d_3=d_4=0 \ . 
\label{full_prop_SO3}
\eeq
The SD equations (\ref{SDc}), (\ref{SDd}) are rewritten 
as 
\beqa
\frac{1}{c_{11}} &=&  
t_1 -2\frac{\del}{\del c_{11}}[G(c, d)]_{{\rm SO(3)}} \ , \nonumber \\
 \frac{1}{\tilde{c}} &=&  
\tilde{t} -\frac23\frac{\del}{\del \tilde{c}}[G(c, d)]_{{\rm SO(3)}} \ , 
\nonumber \\
 \frac{1}{d_1} &=&  
u_1 +\frac{1}{2r}\frac{\del}{\del d_1}[G(c, d)]_{{\rm SO(3)}} \ . 
\eeqa
Here and henceforth the symbol
$[ \ \cdot \ ]_{{\rm SO}(n)}$ implies
that the number of independent variables is already reduced
by the SO($n$) Ansatz.
The differential equations (\ref{diffc}), (\ref{diffd}) become 
\beqa
2\frac{\del}{\del t_1} [{\cal F}(t,u)]_{{\rm SO(3)}} &=& c_{11} \ , \nonumber \\
 \frac23 \frac{\del}{\del \tilde{t}} 
[{\cal F}(t,u)]_{{\rm SO(3)}} &=& \tilde{c} \ , \nonumber \\
 -\frac{1}{2r} \frac{\del}{\del u_1} 
[{\cal F}(t,u)]_{{\rm SO(3)}} &=& d_1 \ .
\eeqa

\subsubsection{SD equations for the SO(2) Ansatz}

The full propagators take the following form.
\beq
c_{\mu\nu}=\left(\begin{array}{cccc} 
c_{11} & c_{12}    &           &         \\
c_{12} & c_{11}    &           &         \\
       &           & \tilde{c} &         \\
       &           &           & \tilde{c} \end{array} 
\right) \ , \qquad d_1=d_2, \quad d_3=d_4=0 \ .  
\label{full_prop_SO2}
\eeq
The SD equations (\ref{SDc}), (\ref{SDd}) are rewritten 
as 
\beqa
\frac{c_{11}}{(c_{11})^2-(c_{12})^2} &=& 
    t_1 -\frac{\del}{\del c_{11}}[G(c, d)]_{{\rm SO(2)}} \ , \nonumber \\
 \frac{c_{12}}{(c_{11})^2-(c_{12})^2} &=& 
     \frac{\del}{\del c_{12}}[G(c, d)]_{{\rm SO(2)}} \ , \nonumber \\
 \frac{1}{\tilde{c}} &=&  
\tilde{t} -\frac{\del}{\del \tilde{c}}[G(c, d)]_{{\rm SO(2)}} \ , 
\nonumber \\
 \frac{1}{d_1} &=&  
2u_1 +\frac{1}{2r}\frac{\del}{\del d_1}[G(c, d)]_{{\rm SO(2)}} \ . 
\eeqa
The differential equations (\ref{diffc}), (\ref{diffd}) become 
\beqa
 \frac{\del}{\del t_1} [{\cal F}(t,u)]_{{\rm SO(2)}} &=& c_{11} \ , \nonumber \\
 \frac{\del}{\del \tilde{t}} 
[{\cal F}(t,u)]_{{\rm SO(2)}} &=& \tilde{c} \ , \nonumber \\
 -\frac{1}{4r} \frac{\del}{\del u_1} [{\cal F}(t,u)]_{{\rm SO(2)}} &=& d_1 \ .
\eeqa



\end{document}